\documentclass[aps,pra,showpacs,twocolumn,amsmath,amssymb,superscriptaddress]{revtex4-1}
\usepackage{graphicx}
\usepackage{dcolumn}
\usepackage{bm}
\usepackage{units}
\usepackage{upgreek}
\usepackage{ucs}
\usepackage{color}
\usepackage{cancel}
\usepackage{hyperref}
\usepackage[all]{hypcap}
\preprint{APS/123-QED}

\usepackage{braket}

\newcommand{\bbR}{\mathbb{R}}

\newcommand{\ud}{\mathrm{d}}
\newcommand{\ic}{\mathrm{i}}
\newcommand{\im}{\mathrm{ Im }\ }

\newcommand{\zZ}{\mathbb{Z}}
\newcommand{\sgn}{\mathrm{sgn}}
\newcommand{\var}{\mathrm{var}}

\begin{document}

\title{Steering random walks with kicked ultracold atoms}
\author{Marcel Wei{\ss}}
\affiliation{ITP, Heidelberg University, Philosophenweg 12, 69120 Heidelberg, Germany}
\author{Caspar Groiseau}
\affiliation{ITP, Heidelberg University, Philosophenweg 12, 69120 Heidelberg, Germany}
\author{W. K. Lam}
\affiliation{Department of Physics, Oklahoma State University, Stillwater, Oklahoma 74078-3072, USA}
\author{Raffaella Burioni}
\affiliation{DiFeST, Universit\`{a} di Parma, Via G. P. Usberti 7/a, 43124 Parma, Italy}
\affiliation{INFN, Sezione di Milano Bicocca, Gruppo Collegato di Parma, Italy}
\author{Alessandro Vezzani}
\affiliation{DiFeST, Universit\`{a} di Parma, Via G. P. Usberti 7/a, 43124 Parma, Italy}
\affiliation{S3, CNR Istituto di Nanoscienze, Via Campi, 213A 41125 Modena, Italy}
\author{Gil S. Summy}
\affiliation{Department of Physics, Oklahoma State University, Stillwater, Oklahoma 74078-3072, USA}
\author{Sandro Wimberger}
\affiliation{DiFeST, Universit\`{a} di Parma, Via G. P. Usberti 7/a, 43124 Parma, Italy}
\affiliation{INFN, Sezione di Milano Bicocca, Gruppo Collegato di Parma, Italy}
\affiliation{ITP, Heidelberg University, Philosophenweg 12, 69120 Heidelberg, Germany}
\email{s.wimberger@itp.uni-heidelberg.de}

\begin{abstract}
A kicking sequence of the atom optics kicked rotor at quantum resonance can be interpreted as a
quantum random walk in momentum space. We show how to steer such a random walk by applying
a random sequence of intensities and phases of the kicking lattice chosen according to a probability 
distribution. This distribution converts on average into the final momentum distribution of the kicked atoms. In particular, it is
shown that a power-law distribution for the kicking strengths results in a L\'evy walk in momentum space 
and in a power-law with the same exponent in the averaged momentum distribution. Furthermore, we investigate 
the stability of our predictions in the context of a realistic experiment with Bose-Einstein condensates.
\end{abstract}



\maketitle

\section{Introduction} 
\label{sec:1}

In many contexts of statistical descriptions in physics and mathematics, random walk models are very useful for qualitative 
and quantitative analysis \cite{randomwalk}. Random walks are one of the simplest stochastic
process and they represent the basic model of diffusion phenomena and non-deterministic motion. When the steps for the diffusing particle
have finite variance and are uncorrelated, the evolutions obey Gaussian statistics and standard diffusion is observed.

In transport and diffusion in complex systems, the basic hypotheses underlying the laws of ordinary Brownian motion can often be violated. In
particular,  the lengths of the steps taken by the diffusing particles can have large fluctuations, and they can
follow a probability distribution with heavy power-law tails, obeying a generalized central limit theorem \cite{econ,probability_theory}.
Examples of the so-called L\'evy \cite{levy2015} random-walk processes are observed in classic transport in complex
materials \cite{wiersma2008,pr2,wiersma2012,wiersma2013,pr1} and in many interdisciplinary contexts in biology,
ecology and economics \cite{econ,malkiel,levy2015}, making these processes a paradigm of transport and
non deterministic evolution in the presence of large deviations.


The concept of a classical random walk can be translated into quantum random walks \cite{PhysRevA.48.1687} using the entanglement between different degrees of freedom, e.g., between the spatial walk variable and a system intrinsic quantity such as spin \cite{kempe2003}. 
An interesting question is how can a quantum walk turn into a classical one and vice versa. 
We address this question for a random walk in the momentum space of kicked cold atoms. We will show how momentum
distributions may be steered almost at will, realizing Gaussian diffusion and more complex $\alpha$-stable distributions \cite{probability_theory}.
The idea of implementing random processes with scale-free behavior with cold atoms in optical lattices goes back to 
pioneering works from the quantum optics community, see, e.g., \cite{zoller1996} which showed that, interestingly,
classical random processes with power-law correlations can be mapped, under certain circumstances, onto a 
deterministic quantum problem (see also \cite{pa2013} for a more recent proposal in this context). 

In particular, it was proposed to exploit the quantum kicked rotor dynamics  
to realize directed quantum walks \cite{PhysRevA.73.013401} and classical L\'evy walks by timing noise implying strong decoherence
of the quantum motion \cite{lutz2007}. As in this latter study, our proposal for a kicked atom realization of random walks 
misses the second degree of 
freedom, which would be entangled with the momenta of the atoms. Yet, we will use the choice of a discrete phase of the 
kick potential to steer the walks, in close analogy to the coin degree of freedom in quantum walk theory \cite{kempe2003}, 
which decides on the random direction of single steps of the walk. 
This is reminiscent of first implementations of quantum walks with classical optics where the coin was also no
second degree of freedom but a random selection by optical elements such as beam splitters 
\cite{silberhorn1,silberhorn2}. In contrast to other studies mentioned above, our quantum motion is, in principle, fully 
reversible for a single realization of the walk, and classicality only results from the classical average over many 
realizations. Our results open the route to future investigations of our system, which fully include the second degree
of freedom. This could be done for cold atoms and Bose-Einstein condensates 
using either internal states of the atoms (an effective spin \cite{PhysRevA.73.013401,schlunk1,WB2006}) 
or more than one kick potential with different phases and wavelengths  
to address independently different momentum classes \cite{sadgrove2012}. 

Our paper is organized as follows: Sec. \ref{sec:2} reviews how quantum resonant motion of the kicked rotor can be used
to realize fast ballistic motion, either symmetric or directed in momentum space. Sec. \ref{sec:3} presents our 
central results showing how the input distribution of phases and, more importantly, of kick strengths convert into
the experimentally easily accessible averaged momentum distributions at quantum resonance. Two specific cases are 
discussed in detail: a Gaussian diffusive walk (\ref{sec:phase}), which can be steered from the quantum to the classical regime, and 
heavy-tailed L\'evy walks (\ref{sec:kwalk}). Secs. \ref{sec:4} and \ref{sec:5} underpin that the results of the previous 
section are indeed robust with respect to typical experimental limitations in atom-optics. 
We show this by checking the stability with respect to small detunings from resonance in sec. \ref{sec:4},
and by taking into account additional complications such as a finite window of feasible kick strengths and a finite width
of a Bose-Einstein condensate in the Brilluoin zone of the periodic kick potential 
(i.e. in quasimomentum) in sec. \ref{sec:5}. Sec. \ref{sec:6} concludes the paper.


\section{Quantum walks at quantum resonance}
\label{sec:2}

Experiments on the quantum kicked rotor based on cold or ultracold atoms work with particles moving along a line periodically kicked in time by an optical lattice. Neglecting atom-atom interactions, the quantum dynamics are described by the following Hamiltonian in dimensionless variables (such that $\hbar=1$) \cite{RaizenAdv,SW2011}:
\begin{equation}
  \label{Hkickr}
  \hat{H}(\hat{x},\hat{p},t)\;=\;\frac{\hat{p}^2}{2} + k\cos( \hat{x}) \sum_{T\in \zZ} \delta(t-T\tau)\ .
\end{equation}
The kick period is $\tau$, the kick strength is $k$, and  $T$ is a discrete time variable that counts the number of kicks.  The periodic potential implies conservation of quasimomentum $\beta$. With the chosen units, $\beta$  can take on allowed values between $0$ and $1$.  
Using Bloch theory, the atom dynamics  from immediately before the $(T-1)$-th kick to immediately before the next $T$-th kick is then described by the one-cycle Floquet operator \cite{SW2011}:
\begin{equation}
  \hat{\cal U}_{\beta,k}(T)\;=\; e^{-\ic \tau(\hat{\cal N}+ \beta)^2/2} \; e^{-\ic k \cos(\hat{\theta})} \ ,
 \label{eq:Flqt}
\end{equation}
where $\hat{\cal N}=- \ic \frac{\ud}{\ud\theta}$ is the (angular) momentum operator with periodic boundary conditions.
In its usual realization with fixed kick strength, the full evolution over $T$ kicks is thus described by
\begin{equation}
  \label{U_tot}
  \hat{\cal
     U}_{\beta}^{T}\;\equiv \;\hat{\cal
     U}_{\beta,k}(T)\; \hat{\cal U}_{\beta,k}(T-1)
   \dots \hat{\cal U}_{\beta, k}(2)\; \hat{\cal U}_{\beta,k}(1)  \,.
\end{equation}
 
A series of experimental investigations has looked at the so-called quantum resonant motion of the quantum kicked rotor \cite{shepelyansky1,Izr1990} over the last decade, see e.g. \cite{SW2011} and references therein. The interest in the resonant dynamics is mainly motivated by the type of ballistic motion with fast acceleration which can be obtained in this particular parameter regime. We restrict our discussion here to the main quantum resonances occurring whenever the kick period $\tau$ is not only commensurate to $2\pi$ but an integer multiple of it, i.e. $\tau=2\pi\ell$, $\ell$ integer. Then, for specially chosen values of quasi-momentum, e.g. for $\beta=1/2$ at $\ell =1$ and $\beta=0, 1/2$  at $\ell = 2$, the energy of the $\beta$-rotor asymptotically increases quadratically with the kick counter $T$ \cite{Izr1990,WGF2003}.  From the theoretical point of view, in contrast to general values of the period, the main quantum resonances are accessible to analytical investigation \cite{Izr1990,WGF2003,FGR2003}. Using the pseudoclassical approximation theory developed in \cite{WGF2003,FGR2003,FGR2006} and reviewed in \cite{SW2011}, this remains true also for small detunings from resonant kick periods. 

\begin{figure}[t!]
\includegraphics[width=\linewidth]{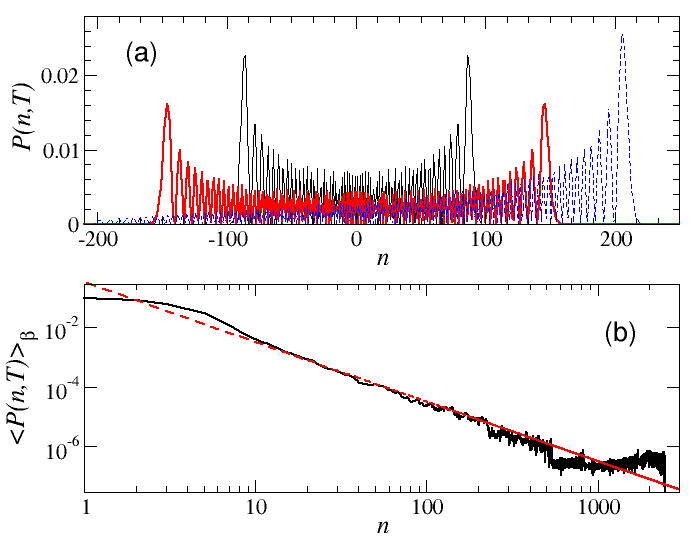}
\caption{(color online). (a) Ballistic motion at quantum resonance for a zero momentum plane wave initial state and $k=3$, either symmetric around $n=0$ ($T=30$ with thin black solid line and $T=50$ thick red/grey solid line), or directed at $T=70$ for a superposition initial state, see text (dashed blue line). (b) Momentum distribution for a symmetric walk as seen in (a) but at $k=0.8\pi$, after $T=1000$ kicks averaged over $10^4$ uniformly distributed quasimomenta $\beta \in [0,2\pi)$. The red dashed line shows a fit according to the  asymptotic formula $4k/(\pi^3n^2)$ derived in \cite{WGF2003}.}
\label{fig:1}
\end{figure}

Since the first factor standing for the free momentum evolution in \eqref{eq:Flqt} is identical to one for the main quantum resonances at resonant quasimomentum, we can easily compute the momentum distribution obtained by just applying the second factor alone. The distribution after $T$ kicks of strength $k$ is therefore the same as the distribution after a single kick with strength $kT$. This gives a momentum distribution after $T$ kicks for an initial state which is an eigenstate of momentum $n_0$ \cite{WGF2003}
\begin{equation} 
\label{eq:bessel}
P(n,T|n_0,k) =  J^2_{n-n_0}(kT) \,.
\end{equation}
Here $J_m$ are ordinary Bessel functions with integer index $m$.
The momentum distribution in the case of $k=3$ and $n_0=0$ is plotted in Fig.~\ref{fig:1}(a). We observe two dominant peaks, which move away from $n=0$ at constant acceleration (i.e. we see ballistic motion). This linear increase in peak momentum is in accordance with the quadratic increase in energy occurring at quantum resonance \cite{Izr1990,WGF2003,FGR2003}. Breaking the spatial-temporal symmetries of the kicked rotor, the quantum resonances permit the realization of directed transport as well \cite{PhysRevLett.94.110603,SW2011,sadgrove2007,SadgroveWimberger2009,sad2013,gil2008}. One example is shown by the thin dashed line in Fig.~\ref{fig:1}(a), where the symmetry is broken by the initial state chosen to be a superposition of two momentum eigenstates $\ket{\psi(n,0)}=(\ket{n=0}+e^{i \phi}\ket{n=1})/\sqrt{2}$ with relative phase $\phi=\pi/2$, see \cite{SW2011,SadgroveWimberger2009} for details.

We already see that the quantum resonances of the kicked rotor can be used to steer the evolution of the wave packet in momentum space. The symmetric ballistic peaks of Fig.~\ref{fig:1}(a) are identical to the wave packets obtained from symmetric quantum random walks, as realized with optical pulses \cite{silberhorn1, silberhorn2}. Also, directed walks are possible when breaking the spatial-temporal symmetry by a correct choice of the initial state, just as in random quantum walks \cite{kempe2003}. The random walk picture best applies in momentum space for our kicked system since each kick couples one specific momentum state to its neighbors, while during the free evolution the momenta are unchanged (``free flight"). The number of effectively coupled neighbors depends on the kick strength $k$. Practically, we have just nearest neighbor coupling for $k\sim 1$ (see Eq. \eqref{eq:bessel} and \cite{WGF2003}). In the following, we ask whether we can steer such a walk in momentum space by other means, e.g., by a random choice of kick strength and/or phases in the sinusoidal kicking potential. For this, we re-derive Eq. \eqref{eq:bessel} for these generalized cases in the next section. Introducing random couplings (whose strength changes with time) makes quantum evolutions classical due to decoherence, see e.g. refs. \cite{RaizenAdv,darcy2001,WGF2003,lutz2007,SadgrovePRE,sadgrove2008} in the context of the quantum kicked rotor. Hence, we can also control the ``quantumness" of the random walks in momentum space, not by changing our effective Planck constant, but by classical averaging over different realizations of the random walk. The effective Planck constants used here throughout are close to $h_{\rm eff}=\tau \approx 2\pi \ell$ \cite{Izr1990,WGF2003,FGR2003}, $\ell$ integer, and hence we are always in the deep quantum regime in this respect.

\section{L\'evy walks in momentum space}
\label{sec:3}

Wrong choices of quasimomentum destroy the ballistic motion  at quantum resonance. Typically, the larger the quasimomentum detuning from its resonant values the faster this will occur \cite{WGF2003}. Hence, averaging quantum resonant motion over a distribution of quasimomentum, e.g., taken uniformly in the unit interval for simplicity, turns the averaged momentum distribution into an average over peaks which stop at different stages in momentum. The asymptotic result for $T\to \infty$ is then a (coarse-grained) momentum distribution in the integers $n$ which decays like a power-law $\sim n^{-2}$. This result was derived in \cite{WGF2003} and rediscovered in \cite{romanelli2010} with some generalizations. Fig.~\ref{fig:1}(b) shows realistic numerical simulations at $k=0.8\pi$ and $T=1000$ which confirm the prediction. Therefore, averaging over various realizations of our random walk naturally leads to a momentum distribution with tails behaving like a Cauchy or Lorentzian distribution. 

Here, we are interested in how to steer a random walk in momentum space by choosing the phases of the kick potential and/or the kick strength randomly from kick to kick. While for a fixed realization, i.e. fixing the series $\{k_t,\phi_t\}_{t=1,\ldots,T}$, the evolution is fully coherent and quantum mechanical, averaging over many realizations, as above for various quasimomenta, makes the result incoherent since the statistical average is a classical one. While this may seem to be a disadvantage, we shall see that this allows us to realize and implement in a realistic experiment with ultracold atoms, essentially any type of random walk we want. All that is required is to choose the appropriate distributions for our input parameters $\{k_t,\phi_t\}_{t=1,\ldots,T}$.

At quantum resonance with $\tau=2\pi\ell$, $\ell$ integer, and ignoring an $n$-independent phase, the Floquet operator simplifies to \cite{WGF2003}
\begin{equation}
\hat{\cal U}_{\beta,k_t}(T) = e^{- i \xi \hat{N}} \, e^{- i k_t \cos(\hat{\theta}+\phi_t)}
\label{eq:floquet_operator}
\end{equation}
with $\xi = \pi \ell (2\beta \pm 1) \operatorname{mod}(2\pi)$ and $\beta$ the quasimomentum. 
The state after $T$ kicks is given by applying consecutively the operator \eqref{eq:floquet_operator} onto the initial state as shown in Eq. \eqref{U_tot}.
Each of these applications is reversible. For example, a kick with $k_j$ and $\phi_j$ can be exactly reversed by a subsequent kick with $k_{j+1}=k_{j}$ and $\phi_{j+1}=\phi_{j}+\pi$. Hence, coherence is preserved during {\it one} complete kicking sequence. Such a time evolution leads to the following time-dependent wave function
\begin{align}
\Braket{\theta|{\psi_{\beta},T}} 
 &= \langle e^{- i \xi \hat{N}} \, e^{- i k_T \cos(\hat{\theta}-\phi_T)}  \, e^{- i \xi \hat{N}} \, e^{- i k_{T-1} \cos(\hat{\theta}-\phi_{T-1})}  \notag \\   
 & \ldots e^{- i \xi \hat{N}} \, e^{- i k_{1} \cos(\hat{\theta}-\phi_{1})}  \ket{\psi_{\beta},j=0} \notag \\ 
&= e^{- i \operatorname{G}(\theta, \xi, \{k_j\}, \{\phi_j\}, T)} \, \psi_{\beta}(\theta-T\xi,0) 
\label{eq:end_spatial}
\end{align}
with
\begin{align} 
& \operatorname{G}(\theta, \xi, \{k_j\}, \{\phi_j\}, T)  = \sum_{j=1}^{T} k_j\cos(\theta-\phi_j-(j-1)\xi) \notag \\ 
& = Re\left(e^{i \theta} \sum_{j=1}^{T} k_j\,e^{- i (\phi_j+(j-1)\xi)} \right) \notag \\ 
& \equiv Re\left(e^{i \theta} \, |R_{T}|\, e^{i \arg(R_{T})} \right) = |R_{T}| \cos(\theta + \arg(R_{T}))
\end{align}
and
\begin{equation} 
R_{T} = R_{T}(\xi, \{k_j\}, \{\phi_j\}) \equiv \sum_{j=1}^{T} k_j\,e^{- i (\phi_j+(j-1)\xi)}
\label{eq:mitquasi}
\end{equation} 
Fourier transforming \eqref{eq:end_spatial} into the momentum ($n$) representation gives
\begin{align} 
& \Braket{n|{\psi_{\beta},T}}  = 
\frac{1}{\sqrt{2\pi}} \int\limits_{0}^{2\pi} e^{-i n \theta} \Braket{\theta|{\psi_{\beta},T}}\,d\theta \notag \\
                                         & = \frac{1}{\sqrt{2\pi}} \int\limits_{0}^{2\pi} e^{-i n \theta} \, e^{- i |R_{T}| \cos(\theta + \arg(R_{T}))} \, \psi_{\beta}(\theta-T\xi,0)   \,d\theta \notag \\
                                         & =  \frac{e^{i n \arg(R_{T})}}{\sqrt{2\pi}} \int\limits_{0}^{2\pi} \, e^{- i (n \theta'+|R_{T}| \cos(\theta'))} \notag \\  
                                         & \ \ \ \ \ \times \psi_{\beta}(\theta'- \arg(R_{T}) - T\xi,0)   \,d\theta' \,,
\end{align}
where the substitution $\theta' = \theta+\arg(R_{T})$ is used. If the initial state of the atom is a plane wave with fixed momentum $p_0=n_0+\beta_0$, then $\xi$ takes the constant value $\xi_0 = \pi l (2\beta_0 \pm 1)\operatorname{mod}(2\pi)$. For the plane wave initial state
\begin{equation}
 \label{eq:initial_wave}
\psi_{\beta_0}(\theta,j\!=\!0) = \frac{1}{\sqrt{2\pi}} \, e^{i n_{0} \theta}
\end{equation}
 and neglecting all global phase factors that will cancel, the following probability distribution is obtained
\begin{align} 
& P(n,T|n_0,\beta_0, \{k_j\}, \{\phi_j\}) = \left| \Braket{n|{\psi_{\beta},T}} \right|^2  \notag \\
                                                         & = \left|  \frac{1}{2\pi} \int\limits_{0}^{2\pi} e^{-i (n-n_{0}) \theta} \, e^{- i |R_{T}| \cos(\theta)}\,d\theta \right|^2 \notag \\
                                                         & = \left|  i^{n-n_{0}} J_{n-n_{0}}(- |R_{T}|) \right|^2 = J^2_{n-n_{0}}(|R_{T}|)\,.
                                                         \label{eq:final}
\end{align}
In the second step, the definition of the Bessel function of first kind is used \cite{AS72}:
\begin{equation}
J_{n}(z) = \frac{1}{2\pi}\int\limits_{\alpha}^{\alpha + 2\pi}e^{-i n \theta} \,e^{ i z cos(\theta)} \,d\theta
\label{eq:besselF}
\end{equation}
together with $J_{n}(-z)=(-1)^n J_{n}(z)$ for integers $n$ \cite{AS72}.

For simplicity, we first focus on the case of resonant quasimomenta, i.e. where $\xi =0$, and refer to section \ref{sec:quasi} for the case of different quasimomenta. Then the argument function is
\begin{equation} 
R_{T} = R_{T}(\{k_j\}, \{\phi_j\}) = \sum_{j=1}^{T} k_j\,e^{- i \phi_j} \,,
\label{eq:RT}
\end{equation} 
which corresponds to a random walk in the complex Argand plane for each realization of the parameters $\{k_j\}_{j=1,...,T}$ and $\{\phi_j\}_{j=1,...,T}$. The length of each step is given by $k_j$ and the direction by the angles $\phi_j$, $|R_T|$ is the displacement from the origin after $T$ steps. Hence, $\rho(|R_T|)$ is the distribution of end displacements for the underlying walk. It is now easy to see that the momentum distribution after a kick sequence with parameters $\{k_j\}_{j=1,...,T}$ and $\{\phi_j\}_{j=1,...,T}$ is equivalent to a distribution that would be obtained by a single kick with effective strength $k_{\rm eff}=|R_{T}|$ (and $\phi_j=0$ for all $j$), or by a sequence of $T_{\rm eff}=|R_{T}|$ effective kicks with strength $k=1$ ($\phi_j=0$ for all $j$). This generalizes the discussion around Eq. \eqref{eq:bessel}, which is regained for $R_T=kT$ at $k=const.$ and $\phi=0$.

In short-hand notation we may now write
\begin{equation} 
\overline{P}(n,T|n_0)  =\int\limits_0^{\infty} d|R_T| \, \rho(|R_T|) \, J^2_{n-n_0}( |R_T|)\,,
\label{eq:Pmean}
\end{equation}
where the distribution $\rho(|R_T|)$ represents the individual averages over the distributions $\rho_{k}(k_j)$ and $\rho_{\phi}(\phi_j)$ assumed to be independent and identically distributed with given probability distributions for the kick strength and phase parts respectively. It is exactly the average from Eq. \eqref{eq:Pmean} in which we are mainly interested since it allows us to steer the random walk in momentum space by appropriate manipulation of $\rho(|R_T|)$.

Focusing on an initial state of the atom with momentum $n_0=0$, the square of the Bessel function $J^2_n(x)$ in Eq. \eqref{eq:Pmean} can be roughly approximated for positive $x$ and large $|n|$ as a Dirac delta function
\begin{equation} 
J^2_n(x) \approx C \, \delta(x-|n|)\,,
\label{eq:bessel_delta_approx}
\end{equation}
where $C$ is an appropriate proportionality constant. It is this property that is responsible for the ballistic ``horns'' seen in Fig. \ref{fig:1}(a). A numerical test and further justification for this approximation can be found in \cite{weiss2014}. Within this approximation, the expression for the averaged momentum distribution in Eq. \eqref{eq:Pmean} simplifies for large $|n|$ to
\begin{eqnarray} 
\overline{P}(n,T|n_0\!=\!0)  &\approx \int\limits_0^{\infty} d|R_T| \, \rho(|R_T|) \, C \,\delta(|R_T|-|n|) \nonumber 
\\ &= C \, \rho(|R_T|\!=\!|n|) \,.
\label{eq:P_final}
\end{eqnarray}
Hence, the large scale behavior of the mean momentum is equivalent to that of the distribution $\rho(|R_T|)$. Consequently, by choosing appropriate distributions $\rho_{k}(k_j)$ and $\rho_{\phi}(\phi_j)$, $\rho(|R_T|)$ and hence the large scale behavior of $\overline{P}(n,T|n_0\!=\!0)$ can be steered in the desired way. 

\subsection{Gaussian random walk}
\label{sec:phase}

\begin{figure}[b]
\includegraphics[width=\linewidth]{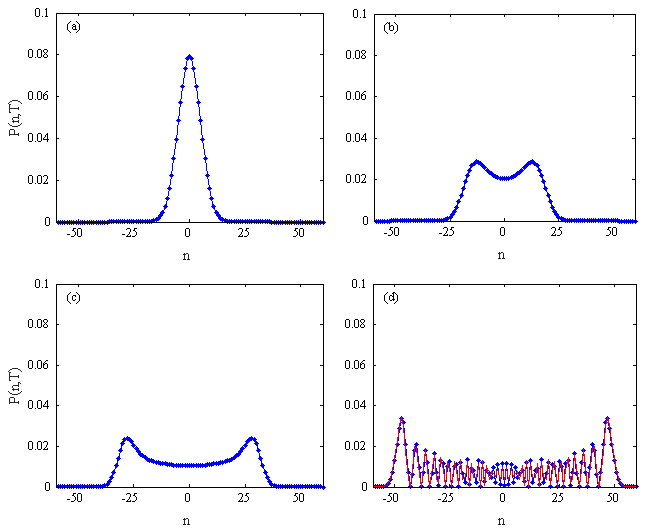}
\caption{(color online). Classical (a) to quantum (d) transition for the random walk in the complex plane (for $T=50$ and $k=1$). In $(a)$ the $\phi_j$s are distributed according to a uniform distribution in $[0,2\pi)$, in $(b)-(d)$ the $\phi_j$s are distributed according to a normal distribution with mean zero and standard deviation $\sigma=1.5$, 1, and 0.1, respectively. In $(d)$ we superimpose the deterministic result for a fixed phase ($\phi=0$) which is identical to the random phase result for small $\sigma$, apart from deviations in the very center at small $|n|<5$.}
\label{fig:2}
\end{figure}

The simplest case of a random walk is obtained when only the phase of the kicking potential $\phi$ is taken from a random distribution at fixed $k$. Since the phase is defined between zero and $2\pi$ there is not much choice for non-standard, e.g., heavy tailed distributions. We choose a Gaussian distribution, whose two limits are a delta function centered at some fixed value of the phase and a uniform distribution for a standard deviation much larger than $2\pi$. The former case obviously reduces to the deterministic walk reviewed in Fig.~\ref{fig:1}(a). This limit is highlighted in Fig. \ref{fig:2}(d). The latter case was analyzed in great detail in ref. \cite{WGF2003}, for random jumps in the quasimomentum. In a changed frame of reference this is identical to jumps of the phase, the only difference lying in the fact that commutative shifts of quasimomentum induce additional shifts of momentum itself (an additional heating effect not present here). We therefore can be brief in translating the phase walk in the complex plane into the language of a momentum walk. 

We may interpret the end position of the complex valued function $R_T=(X,Y)$ as a vector in an Argand diagram. The evolution of $R_T$ corresponds to a diffusion in the complex plane, whose details are determined by the phase distribution. In all non-pathological cases, according to the central limit theorem, the distribution of $(X,Y)$ for large $T$ is hence given approximately by a two-dimensional normal distribution:
\begin{equation}
f\left(R_T=(X,Y)\right) = \frac{1}{2 \pi  \sigma_X \sigma_Y} e^{-\frac{1}{2}\left[\frac{(X-\mu_X)^2}{\sigma_X^2} + \frac{(Y-\mu_Y)^2}{\sigma_Y^2}\right] }\,,
\label{eq:2d_normal}
\end{equation}
where $\mu_i$ and $\sigma_i$ ($i=X,Y$) are the mean and standard deviation of $X$ and $Y$, respectively. We obtain for the distribution of the random variable $R_T$ that the mean $\operatorname{E}(R_{T}) = \mu_X + i \, \mu_Y$ and the variance $\operatorname{Var}(R_{T}) = \sigma_X^2 + \sigma_Y^2$. By considering the $\phi_j$s as  taken from a symmetric distribution centered around zero, $\operatorname{E}(R_{T})$ will be real and consequently $\mu_Y=0$. The smaller the width of such a distribution for the $\phi_j$s, the more the random walk will be directed and $\operatorname{E}(R_{T})=\mu_X$ shifted along the positive real axis away from the origin. 

For a uniform distribution in $[0,2\pi)$ for the $\phi_j$s, it is: $\operatorname{E}(R_{T})=\mu_X=0$ and $\sigma_X^2=\sigma_Y^2= T k^2 /2 $, and we obtain
\begin{equation}
f(R_T) = \frac{1}{\pi T k^2} e^{\left(-\frac{X^2+Y^2}{T k^2} \right)} = \frac{1}{\pi T k^2} e^{\left(-\frac{|R_T|^2}{T k^2} \right) }\,.
\end{equation}
Multiplying by $2\pi$ times the radius $|R_T|$, leads to the following distribution for the end displacement $|R_T|$
\begin{equation}
\rho(|R_T|)= \frac{2}{T k^2} \exp\left(- \frac{|R_T|^2}{T k^2} \right) |R_T| \,.
\label{eq:end_displacement_uniform}
\end{equation}
This distribution has a peak close to zero and shows a normal behavior. For non-uniform distributions of $\phi_j$s, $ \sigma_X \neq \sigma_Y$. Hence, Eq. \eqref{eq:2d_normal} cannot be rewritten in a closed analytical form for the distribution $\rho(|R_T|)$. The precise form of the phase distribution, $\rho_{\phi}(\phi_j)$, can be used to steer the final distribution $\rho(|R_T|)$, e.g. away from zero with $\mu_X \neq 0$. 
Because of Eq. \eqref{eq:P_final}, the averaged momentum distribution for large $|n|$ obeys $\overline{P}(n,T|n_0\!=\!0) \propto \rho(|R_T|)$, so that the final momentum distribution  follows the distribution of $\rho(|R_T|)$.
As seen in Fig. \ref{fig:2}, we may steer the random walk from quantum, panel (d), to classical, panel (a), by controlling the widths $\sigma_{X}=\sigma_Y$ of the phase distribution.

\subsection{Power-law walks in momentum space}
\label{sec:kwalk}

We have seen above that random phase shifts alone cannot lead to more interesting non-Gaussian distributions. In consequence, we must include the possibility of randomizing the kicking strength $k$ in Eq.~\eqref{eq:RT} as well. Obviously, a Gaussian distribution with fixed mean and standard deviation would have the same consequences as the case just discussed in the previous subsection (with the only difference being a change in the final Gaussian momentum distribution). 

However, more interesting choices are possible since $k$ is not bounded from above, just from below at $k=0$. For simplicity, we can interchange a distribution of just positive values of $k$ with a symmetric distributions around zero with. This is possible since the phase $\phi_j$ may be chosen from a discrete uniform distribution, which only takes two values with equal probability $p=1/2$:
\begin{equation}
\rho_{\phi}(\phi_j) = \begin{cases}0 & p\!=\!\frac{1}{2} \\ \pi & p\!=\!\frac{1}{2} \end{cases} \,.
\label{eq:phi_discrete}
\end{equation}
From the form of the kick evolution operator, Eq. \eqref{eq:Flqt}, it can be seen that a $\pi$ phase shift is equivalent to a sign change in front of $k$. Such a phase shift is experimentally feasible by varying the relative position between the atoms and the kicking standing wave \cite{gil2010,gil2013a,sad2013,hoogerland2013}. Furthermore, asking which of the two phases should be chosen before each kick, is analogous to the coin toss in standard quantum random walk algorithms, see e.g. \cite{kempe2003}. 

In the following, we focus on so called $\alpha$-stable distributions $S(\alpha,\gamma,\mu)$ for the probability distribution of the kick strength. These have several advantages, the most important one is that they are closed under convolution for a fixed value of the parameter $\alpha$, in the sense that the sum of $N$ independent and identically distributed random variables $X_j \sim S(\alpha,\gamma,\mu)$ is again distributed with $\sum_{j=1}^{N } X_j \sim S(\alpha,N\gamma,N\mu)$ \cite{probability_theory}. Here the first parameter $\alpha \in (0,2]$ is called the characteristic exponent, it describes the tail of the distribution. $\gamma > 0$ is a scale parameter characterizing the width of the central part of the distribution and $\mu \in \bbR$ determines the position of its center \cite{probability_theory}. For $ \alpha < 2$, the stable distributions have an infinite variance and asymptotically decrease as $|x|^{-(1+\alpha)}$ \cite{probability_theory}. Hence, by choosing $\mu\!=\!0$, the resulting distribution $S(\alpha,\gamma,0)$ is a distribution symmetric around zero with power-law tails.

Most of our data is produced now when the $k_j$s obey a Cauchy distribution, as a special case with $\alpha=1$. We formally assume that $k$ can be negative to simplify the argument. Here an explicit form of $S(\alpha,\gamma,\mu=0)$ can be given:
\begin{equation}
\label{eq:cauchy}
\rho_k(k_j) = 
{ 1 \over \pi }  { \gamma \over  \gamma^2 + k_j^2}  
\,.
\end{equation}
Let us also assume for a moment that $\phi=const.$ (including the case of jumping between $0$ and $\pi$ for the realization of effectively negative $k$). From the fact that the kick strengths $k_j$ are Cauchy distributed with $S(1,\gamma,0)$, it follows that the sum $R_T= \exp(-i\phi) \sum_{j=1}^{T} k_j$ is also Cauchy distributed with $R_T \sim S(1,T\gamma,0)$. $|R_T|$ is then the ``length" of the end positions of a one-dimensional L\'evy walk on the real axis. The distribution of end displacements is therefore given by
\begin{equation}
\rho(|R_T|) = { 2 \over \pi } { T\gamma \over (T\gamma)^2 + |R_T|^2}\,. 
\label{eq:end_displacement}
\end{equation}
For the large time asymptotics, it does not matter whether $\phi=const.$ as just assumed (corresponding to an effectively one-dimensional walk) or drawn from a uniform distribution in $[0,2\pi)$ (fully two-dimensional walk in the complex plane), or anything intermediate. For large $|R_T|$, the distribution will always be similar to $S(1,\tilde{\gamma},0)$, with a scale factor $\tilde{\gamma}<T\gamma$ since the optimal spread is achieved exactly for the one-dimensional walk. Thus the tails are power-law distributed in all cases, with the same power-law (i.e. the same $\alpha$) as the kick strength distribution.

\begin{figure}[t!]
        \includegraphics[width=1.1\linewidth]{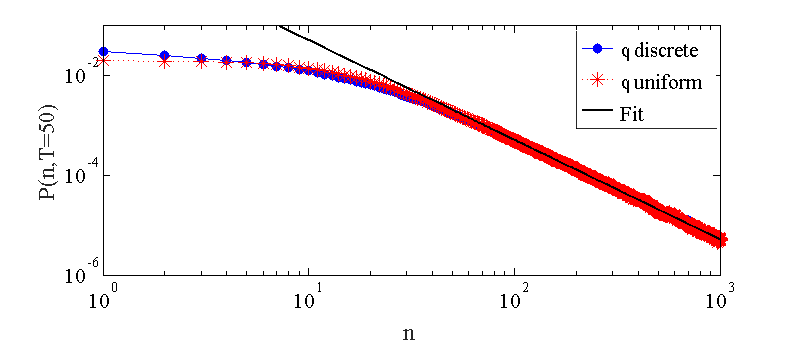}
\caption{(color online). Numerically obtained averaged momentum distributions with $50{,}000$ realizations for the $\phi_j$s and $k_j$s at $T=50$, $n_0=0$ and $\xi=0$. For the $k_j$s a Cauchy distribution with scale parameter $\gamma=0.5$ is used taking into account just positive values of $k$ (see also Eq. \eqref{eq:cauchyK} below), for the $\phi_j$s the discrete distribution  from Eq. \eqref{eq:phi_discrete} (blue line with dots) and a uniform distribution (red dotted line with asterisks) is considered. The thick black line represents a power law with exponent $2$ (or an $\alpha$ parameter $\alpha=1$) fitted to the data.}
\label{fig:3}
\end{figure}

Figure~\ref{fig:3} shows in a log-log plot the results of numerical simulations carried out at quantum resonance for $50{,}000$ realizations of the walk for with different series $\{\phi_j\}$s and $\{k_j\}$s at $T=50$, $n_0=0$ and $\xi=0$ (resonant quasimomentum). The plot shows a comparison of the averaged momentum distribution obtained by considering for the $k_j$s a Cauchy distribution with only positive values, scale parameter $\gamma=0.5$, and for the $\phi_j$s either $0$ or $\pi$ (blue line with dots) or a uniform distribution in $[0,2\pi)$ (red line with asterisks). Recall the possibility of $\phi_j > \pi$ effectively corresponds to negative kicking strength. Both results are identical in the tails, showing a power law with exponent $\alpha=1$ at large momenta $|n|>30$. 

To summarize by choosing the kicking strengths from an appropriate power-law distribution $\rho_k(k_j)$, a L\'evy walk can be realized in the momentum space of the atoms with the same asymptotic power law as the one of the input distribution $\rho_k(k_j)$. The momentum distribution is the quantity typically measured in an atom-optics experiment \cite{RaizenAdv,SW2011}.

\section{Stability with respect to detunings: $\epsilon$-classical analysis}
\label{sec:4}

Quantum mechanically, there is no general analytic solution outside the quantum resonance condition for the quantum kicked rotor dynamics, and we may refer only to numerical simulations, as we will do below. However, if we detune the kick period $\tau$ just slightly from the resonance conditions, we may still be able to estimate the true quantum motion using the pseudoclassical model introduced in \cite{WGF2003,FGR2003}. In this model, the absolute value of the detuning $|\epsilon|$ plays the role of Planck's constant, and hence the theory has a semiclassical limit at exact quantum resonance.  The $\epsilon$-classical dynamics are described by the following discrete map which relates the variables $I$ and $\theta$ from the $j$-th kick to the next one:
\begin{equation}
  \label{eq:cl}
    \begin{array}{rcl}
      I_{j+1}&=&I_{j} + \tilde{k}_{j+1}\sin(\theta_{j}+\phi_{j+1}) \\
      \theta_{j+1}&=&\theta_{j}+\sgn(\epsilon)I_{j+1}\;\;\;\textrm{ mod } 2\pi
      \end{array}
      \,.
\end{equation}
Here, ${I}={J}+\sgn(\epsilon)[\pi \ell+\tau\beta]$ and $J=|\epsilon| p$ is the rescaled momentum. Similarly, a new effective kick strength is defined by $\tilde{k}_j=|\epsilon| k_j$, which is multiplied by the small detuning in this model, making the classical phase space nearly integrable. Provided that the effective kick strength $\tilde{k}$ remains small, the
map \eqref{eq:cl} yields a good approximation of the true quantum motion \cite{FGR2003,WGF2003}. The momentum distribution studied in the previous sections corresponds to a distribution of $p(T,\theta_0)=J(I_T,\theta_0)/|\epsilon|$ for the initial ensemble of $\theta_0$ (uniformly distributed in $[0,2\pi)$ for modeling a plane wave initial condition with fixed momentum) and the realizations of the random variables $\{\phi_j\}, \{k_j\}$. The momenta $I_T$ after $T$ kicks are given by
\begin{equation}
  \label{eq:cl-1}
I_{T}=I_{0} + \sum_{j=0}^{T-1} \tilde{k}_{j+1}\sin(\theta_{j}+\phi_{j+1})\,.
\end{equation}
To begin with, we look at the case of fixed $k$ and broadly distributed phases, just as in subsection \ref{sec:phase}. Then the single momentum changes are uncorrelated and the motion will be diffusive just as assumed for the deterministic standard map in the chaotic regime \cite{chirikov,shepelyanskyqcorr}. Formally, the momentum increase $\Delta I_T \equiv I_T - I_0$ is related to our distribution function $R_T$ from Eq. \eqref{eq:RT}, by using $\tilde{\phi}_{j+1}\equiv - (\theta_j+\phi_{j+1})$ and rewriting
\begin{equation}
  \label{eq:cl-2}
\Delta I_T = - \im \left( \sum_{j=1}^T \tilde{k}_j e^{-i\tilde{\phi}_j} \right) = - \sum_{j=1}^T \tilde{k}_j \im \left(e^{-i\tilde{\phi}_j} \right) \,.
\end{equation}
The random walk obtained in the variable $I$ is then just a projection of the walk in the complex plane onto the imaginary axis. Retranslating into the original momentum variable $p$, gives a result very similar to the case studied in sections \ref{sec:phase} and \ref{sec:kwalk}, when the phases $\phi_i$ strongly vary in the interval $[0,2\pi)$.

\begin{figure}[t]
\includegraphics[width=0.9\linewidth]{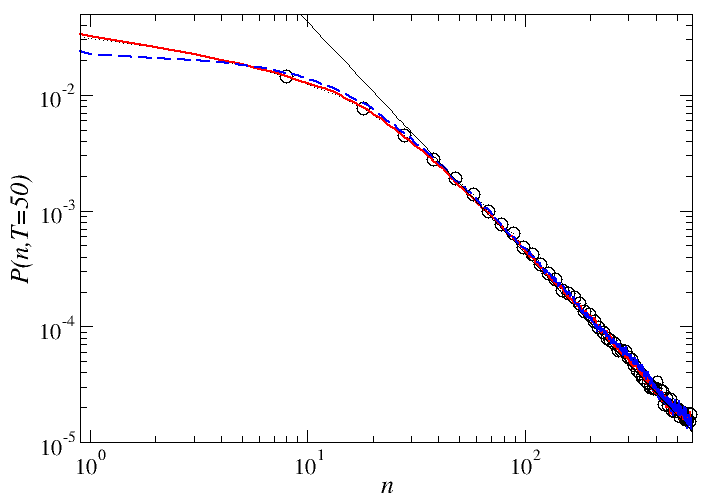}
\caption{(color online). Averaged momentum distributions computed by evolving the quantum map generalizing Eq. \eqref{U_tot} with randomly distributed $k_j$ from a Cauchy distribution (no phase changes). Parameters are $\tau= 4\pi+\epsilon$, $\gamma=0.5$ after $T=50$ kicks, and $20000$ realizations are taken into account. Data are shown for three different detunings $\epsilon = 0.001$ (black dotted line with circles), $0.01$ (red/grey thick solid line), and $0.1$ (blue dashed line). The different curves only differ in the center close to zero momentum while the tails are identical. The case $\epsilon = 0.001$ is not distinguishable from the result at exact quantum resonance, see blue line with circles in Fig. \ref{fig:3}. The black solid line shows a power-law fit with exponent $2$ as expected for the Cauchy case.
 }
\label{fig:4}
\end{figure}

For a fixed phase $\phi$, the momentum increase will dominantly grow by the sum of kick strengths, since we may assume a random reshuffling of the angles due to the random change in $k$. This is similar to the argument used to derive momentum diffusion in the deterministic standard map \cite{chirikov,shepelyanskyqcorr}, where chaos there is substituted here by true randomness.
In other word, we may approximate 
\begin{equation}
  \label{eq:cl-3}
		\sum_{j=1}^T \tilde{k}_j\sin(\theta_{j}+\phi) \sim  \sqrt{\var\left(\sin(\theta_j+\phi)\right)} \sum_{j=1}^T \tilde{k}_j \sim  \frac{1}{\sqrt{2}} \sum_{j=1}^T \tilde{k}_j \,,
\end{equation}
with $\var\left(\sin(\theta_j+\phi)\right) \approx 1/2$ from the approximations $\langle \sin(\theta_{j+1})\sin(\theta_{j})\rangle \approx 0 $ and $\langle \sin(\theta_{j})\sin(\theta_{j})\rangle \approx\int_o^{2\pi} dx\sin^2x =1/2$. Hence, we expect that the walk in momentum space can be steered by the distribution of kick strengths close to quantum resonance making it robust with respect to experimental imperfections. Our expectation, based on the crude approximation from Eq. \eqref{eq:cl-3}, is confirmed by full quantum simulations at finite detunings, which are shown in Fig. \ref{fig:4}, as well as by pseudo-classical simulations based on the above reasoning (not shown here).

\section{Experimental implementation}
\label{sec:5}

We now investigate how the predictions of the previous section could be applied in a state-of-the-art experiment with Bose-Einstein condensates, such as the one successfully running at Oklahoma State University \cite{gil2008,gil2010,gil2010b,gil2012,gil2013a,gil2013b}. We focus, in particular, on possible problems in the implementation of random walks with heavy tail statistics. It is clear that the experimental challenge is two-fold: first, relatively large momenta should be observed, at least over one or two orders of magnitude. Second, the signal in the wings should be large enough to give an experimental signal above the noise level, see e.g. \cite{DarcyPRE,Wimberger2005PRA}. Next, practical problems in the realization of the ideal walk must be considered, e.g., the challenge of rapidly changing the kick strength from kick to kick over an appreciable range of values, or the finite width of a Bose-Einstein condensate in quasimomentum. In the subsequent subsections we will address these and other possible effects.

\subsubsection{Fixing the kicking strength for a sequence}
\label{sec:kickS}

The first issue to address is that it may be easier not to change the kick strength for every kick during a measurement sequence. Therefore, our proposed idea is to fix the kick strength $k_j=k$ for $j=1,...,T$ during a sequence, and to choose for every repetition of the experiment a different $k$ drawn, for example, from a heavy tailed power-law distribution. Finally, an average over all results can be performed to obtain the walk momentum distribution. In such a case, the individual experimental run is no longer a random walk in momentum space since $k$ is fixed. However, if the potential shifts $\phi_j$ can be adjusted precisely during a sequence, a Gaussian walk trajectory, as studied in \ref{sec:phase}, can be realized for each run. We will show now that in this case, the final averaging over many realizations eventually leads to a heavy tailed momentum distribution as well even if $k$ is fixed for each repetition but chosen randomly from an $\alpha$-stable distribution.

\begin{figure}[t!]
        \includegraphics[width=1.05\linewidth]{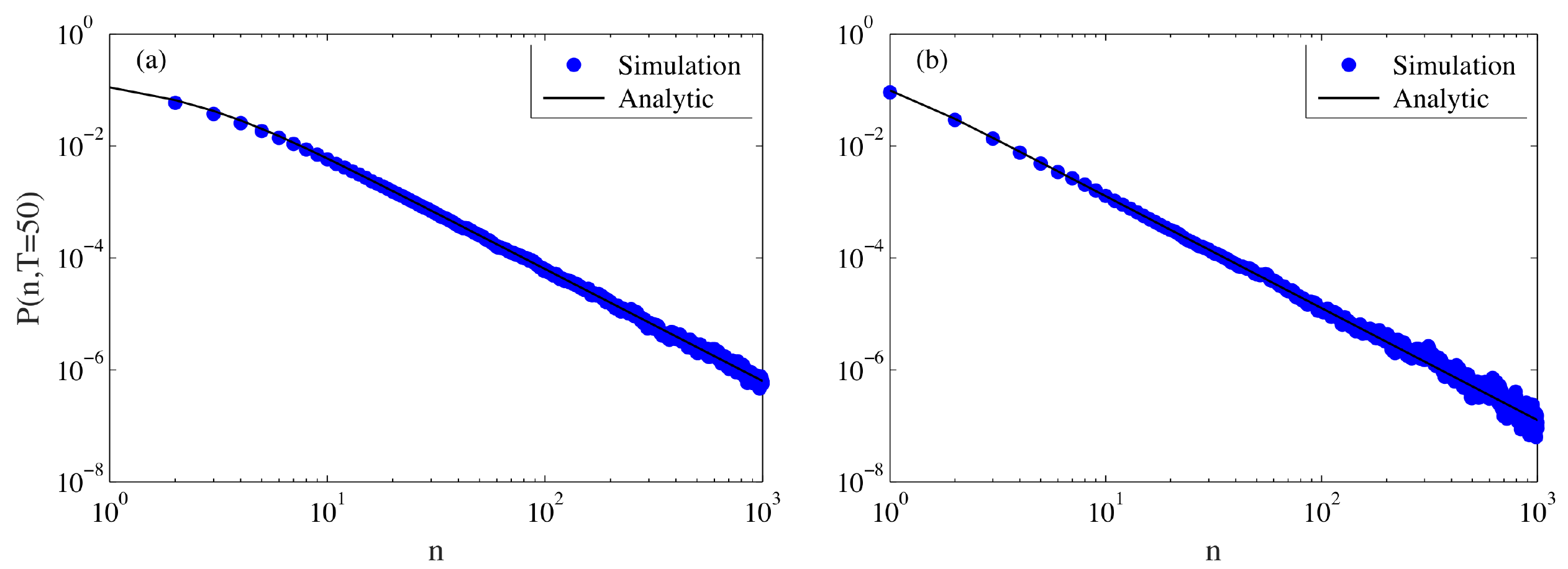}
\caption{(color online) Log-log plots of the numerically obtained averaged momentum distributions (blue symbols) via simulations with $50{,}000$ realizations for the $\phi_j$s and $k$, $T=50$, $n_0=0$ and $\xi=0$. For the $\phi_j$s a uniform distribution is considered, for $k$ the Cauchy distribution Eq. \eqref{eq:cauchyK} with scale parameter $\gamma=0.5$ [panel $(a)$] and $\gamma=0.1$ [panel $(b)$]. Additionally, we show the averaged momentum distribution obtained by computing Eq. \eqref{eq:Pmean} with the derived analytical expressions for $\rho(|R_T|)$  (black solid line) from Eq. \eqref{eq:averageK}.}
\label{fig:5}
\end{figure}

Following Eq. \eqref{eq:Pmean}, at quantum resonance, the mean momentum distribution after $T$ kicks is now explicitly given by 
\begin{eqnarray} 
\overline{P}(n,T|n_0)  =  \int\limits dk\, \rho_{k}(k) \left[  \int\limits d\phi_1\, \rho_{\phi}(\phi_1)  \right. ... \\ \notag
\left. \int\limits d\phi_T\, \rho_{\phi}(\phi_T)\, J^2_{n-n_0} \left( |R_T| \right) \right]
\label{eq:average}
\end{eqnarray}
where only one integral over $k$ needs to be considered. We must now derive an expression for the distribution $\rho(|R_T|)$, with $R_T=\sum_{j=1}^{T} k \,e^{-i \phi_j}$, a random walk in the complex plane with fixed jump length $k$ for each individual repetition of the experiment. We focus on the simplest case in which the $\phi_j$s are chosen from a uniform distribution in $[0,2\pi)$. From section \ref{sec:phase}, it is known that for fixed $k$ and uniformly distributed $\phi_j$s, the distribution of end displacements is given by Eq. \eqref{eq:end_displacement_uniform}, which leads to the following mean displacement
\begin{equation}
\overline{|R_{T}(k)|}  = \int\limits_{0}^{\infty} |R_{T}(k)| \, \rho(|R_{T}(k)|)  \, d|R_{T}(k)| = \dfrac{\sqrt{T \pi}}{2} \, |k| \,.
\end{equation}
The influence of the $\phi_j$s, which determine the displacement of a single sequence with fixed $k$, is now neglected since it is small compared to the possibly large jumps in $k$. Then we can assume that the displacement of a single sequence is given approximately by the mean
\begin{equation} 
|R_{T}(k)| \approx \overline{|R_{T}(k)|} = \dfrac{\sqrt{T \pi}}{2} \, |k| \,.
\end{equation}
Since this mean displacement is linearly dependent on $|k|$, the resulting distribution $\rho(|R_{T}|)$ is again completely dominated by the chosen distribution for $|k|$. To be specific we now consider a ``positive" Cauchy distribution with $k \geq 0$
\begin{equation}
\label{eq:cauchyK}
\rho(k) \ = \  \dfrac{ 2 \gamma}{ \pi ( \gamma^2 + k^2)} \,,
\end{equation}
where the factor two corrects the normalization due to taking into account only positiv values.
By substitution $\rho(k) \, dk \approx  \rho(|R_{T}|) \, d|R_{T}|$ we arrive at
\begin{equation}
\label{eq:averageK}
\rho(|R_{T}|) \approx \dfrac{ 2 \left( \dfrac{\sqrt{T \pi}}{2} \gamma \right)}{ \pi \left( \left(\dfrac{\sqrt{T \pi}}{2}\gamma\right)^2 + |R_{T}|^2\right)}  \,.
\end{equation}
As in section \ref{sec:kwalk}, the distribution of $k$'s from Eq. \eqref{eq:cauchyK} carries over to a Cauchy distribution for the end displacement  $|R_{T}|$, 
although with a smaller width, $\sqrt{T \pi}\gamma/2$ as compared to $T\gamma$. Figure \ref{fig:5} confirms the validity of this approximate result.
The plots also show that, at smaller $\gamma$, the power law starts at lower values of momentum, while at larger $\gamma$ the power-law starts at larger momenta.


\subsubsection{Taking into account the quasimomenta}
\label{sec:quasi}

Up to now, always resonant quasimomenta have been considered, corresponding to  $\xi=0$ in Eq. \eqref{eq:mitquasi}. 
Modern BEC experiments allow one a relatively precise adjustment of the $\beta$ distribution which is approximated to be Gaussian \cite{gil2012,gil2013a}. Sticking to the quantum resonance at $\tau=2\pi$ (i.e. $\ell=1$), we consider the distribution centered around the resonance value $\beta=0.5$:
\begin{equation}
\rho_{\beta}(\beta) =  \frac{1}{\sigma\sqrt{2\pi}} \, e^{ -\frac{(\beta-0.5)^2}{2\sigma^2} } \,,
\end{equation}
with $\sigma= \nu /2/{\sqrt{\ln(2)\,2}}$, where $\nu \approx 0.05$ is the full width at half maximum (FWHM) measured in \cite{gil2013a}. An adequate value for the number of considered $\beta$s is $B=10{,}000$ corresponding to the typical atom number in the experiment \cite{gil2008,gil2010,gil2010b,gil2012,gil2013a,gil2013b,hoogerland2013}. This implies the assumption that each atom has its own quasimomentum, neglecting interactions, and hence we assume that the initial atomic ensemble is an incoherent mixture of plane waves with different but fixed quasimomenta.

Hence, given the probability distribution after a kicking sequence for an atom with fixed $\beta=\beta_0$, $P(n,T|n_0,\beta_0,\{k_j\},\{\phi_j\})$, the ``real'' momentum distribution, under the incoherent approximation, is given by the following classical average
\begin{eqnarray}
& P(n,T|n_0,\{k_j\},\{\phi_j\})  = \nonumber \\ 
& \ \ \ \ \ \int\limits_0^1 d\beta \, \rho_{\beta}(\beta) \, P(n,T|n_0,\beta,\{k_j\},\{\phi_j\})\,.
\end{eqnarray}
At quantum resonance and for a fixed $\beta$, $P(n,T|n_0,\beta,\{k_j\},\{\phi_j\})= J^2_{n-n_{0}} (|R_{T}|)$, see Eq. \eqref{eq:final}, where $R_{T}$ now takes the extended form
\begin{equation} 
R_{T} = R_{T}(\{k_j\}, \{\phi_j\}, \xi(\beta)) = \sum_{j=1}^{T} k_j\,e^{- i (\phi_j+(j-1)\xi)}
\label{eq:R_mod}
\end{equation}
This formula underlines the similarity between $\xi$ (and hence $\beta$) and $\phi_j$ acting both as effective potential displacements. Both have an influence on the direction of the single steps performed in the underlying walk in the complex plane. If we consider a case in which the $\phi_j$s are uniformly distributed in $[0,2\pi)$, the potential is completely randomly shifted in $\theta$ space for every kick. Further shifts arising from the different $\beta$ are therefore averaged out and negligible. Thus, in this special case, the effect of the $\beta$s can be absorbed in the phase average. 

\begin{figure}[t!]
    \begin{center}\vspace{5mm}
        \includegraphics[width=0.9\linewidth]{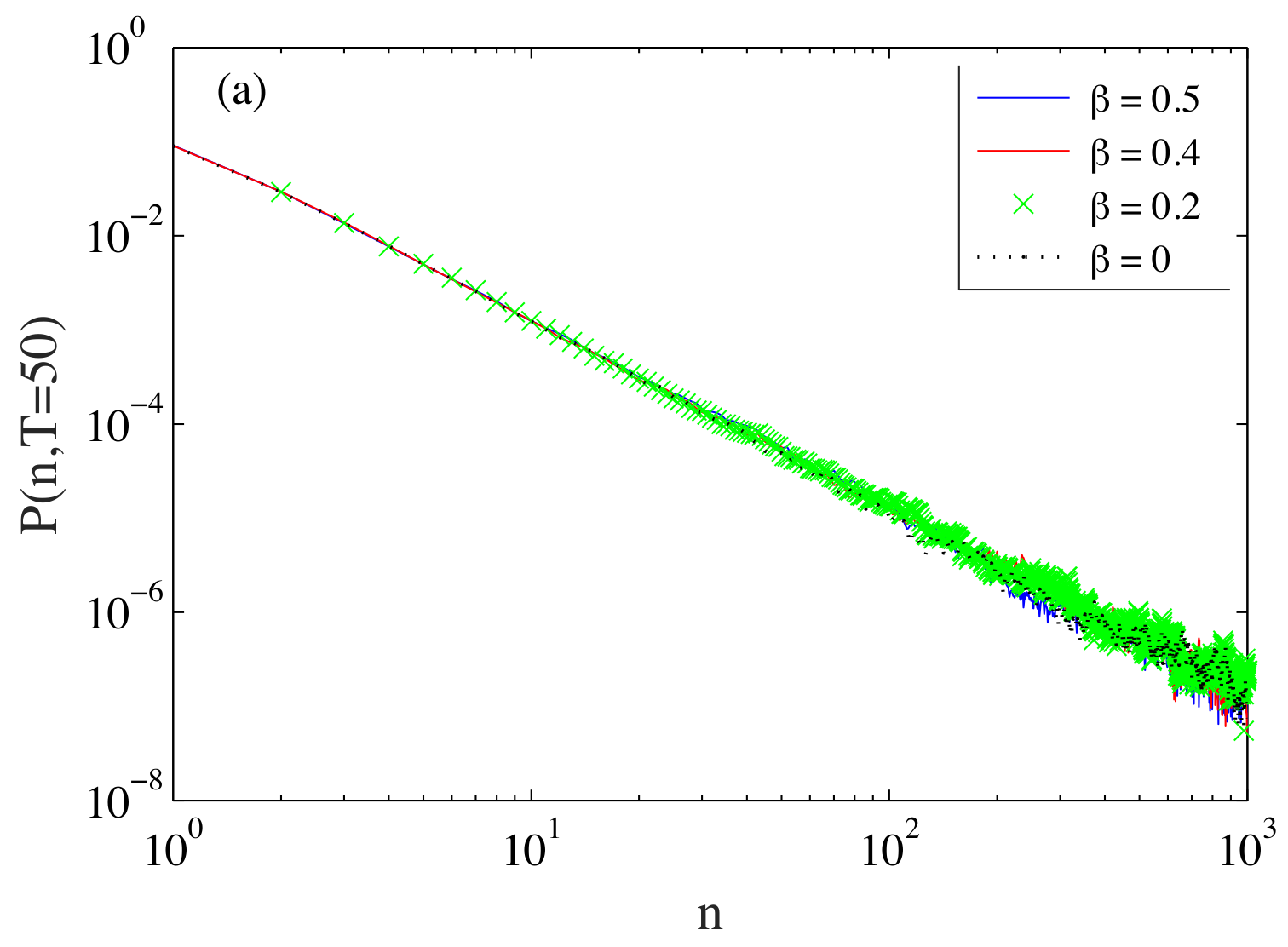}
        \hspace*{3mm}\includegraphics[width=0.885\linewidth]{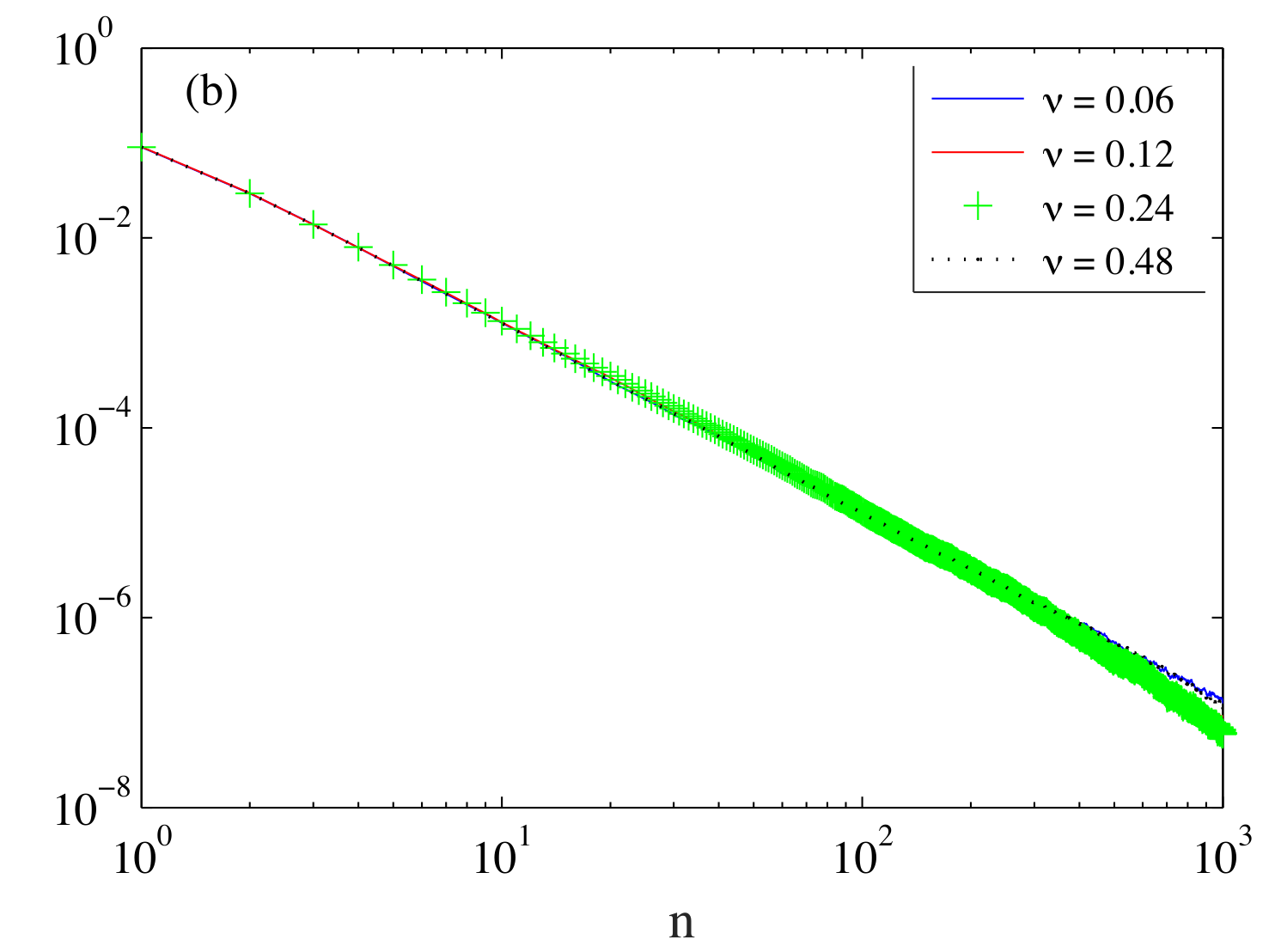}
    \end{center}
\caption{Log-log plots of numerically obtained averaged momentum distributions via simulations which take into account the quasimomenta. We consider the $\phi_j$s uniformly distributed, $k$ fixed for each run (see section \ref{sec:kickS}) and chosen from the Cauchy distribution \eqref{eq:cauchyK} for each repetition; the parameters are $T=50$, $n_0=0$ and $\tau=2 \pi$. In $(a)$ the results are plotted by considering $R=50{,}000$ realizations for the $\phi_j$s and $k$ and the different single values for $\beta$ given in the legend. In $(b)$ the results are plotted by considering $R=25{,}000$ and $B=200$ values for $\beta$ chosen from a normal distribution centered around the resonance value $\beta=0.5$ with different values for the FWHM $\nu$, see legend.}
\label{fig:6}
\end{figure}

Figure \ref{fig:6} shows the results of numerical simulations for which the additional averaging over the $\beta$s is done. We consider the case in which the $\phi_j$s are uniformly distributed and $k$ is fixed and chosen from the Cauchy distribution \eqref{eq:cauchyK} for each repetition. First, for the simulation in panel $(a)$, $50{,}000$ realizations for the $\phi_j$s ($j=1,...,T$) and $k$ are considered at only one non-resonant value for $\beta$, i.e. at $\tau=2\pi$ with $\beta \neq 0.5$. In all cases, one obtains a similar result and even for $\beta=0$, which is the value the furthest from resonance, no significant difference is observed. In panel $(b)$, several normal distributions for $\beta$ are considered with a smaller number of realizations. Again, for all considered values of $\nu$, no significant difference is seen. The differences for higher $|n|$ result from the fact that even $R=25{,}000$ realizations are not  sufficient to obtain always a power law up to the tails at $|n|>10^3$ with significant statistics. In the next sections, simulations with smaller values for $R$ are done, thus giving the opportunity for additional checks with a more realistic number of quasimomenta.

\subsubsection{Limits for the kicking strength and the number of repetitions}
\label{sec:limits}

Another experimental problem is that no infinite range for the kicking strength can be realized, which would be necessary to cover the entire tail of a power-law input distribution. Given the typical laser intensities in the laboratory, one may realize without problems a range of $k$ between about $0.1$ and $5$. 
The lower limit is set by experimental uncertainties, which make it hard to distinguish between $k=0$ and $k=0.1$ kicking signals. Hence, only a small part of the Cauchy distribution for $k \geq 0$ can be considered:
\begin{equation}
\rho_k(k) = \begin{cases} const. \, { 2 \over \pi }  { \gamma \over k^2 + \gamma^2  }  & \text{for } k_{\rm min} \le k \le k_{\rm max} \\  0 & \text{otherwise }  \end{cases}
\label{eq:limits}
\end{equation}
where $const.$ is an appropriate normalization constant. It is suitable to use $\gamma=0.1$ since for such a small scale parameter, the power law sets in early and covers the accessible interval for $k$. Furthermore, since one measurement takes about $30$ seconds, $1{,}000$ repetitions are reasonable.
Because of the expected cutoff at large momenta, it is also sufficient to consider only $1{,}000$ realizations from a statistical perspective. 

\begin{figure}[t!]
        \includegraphics[width=1.05\linewidth]{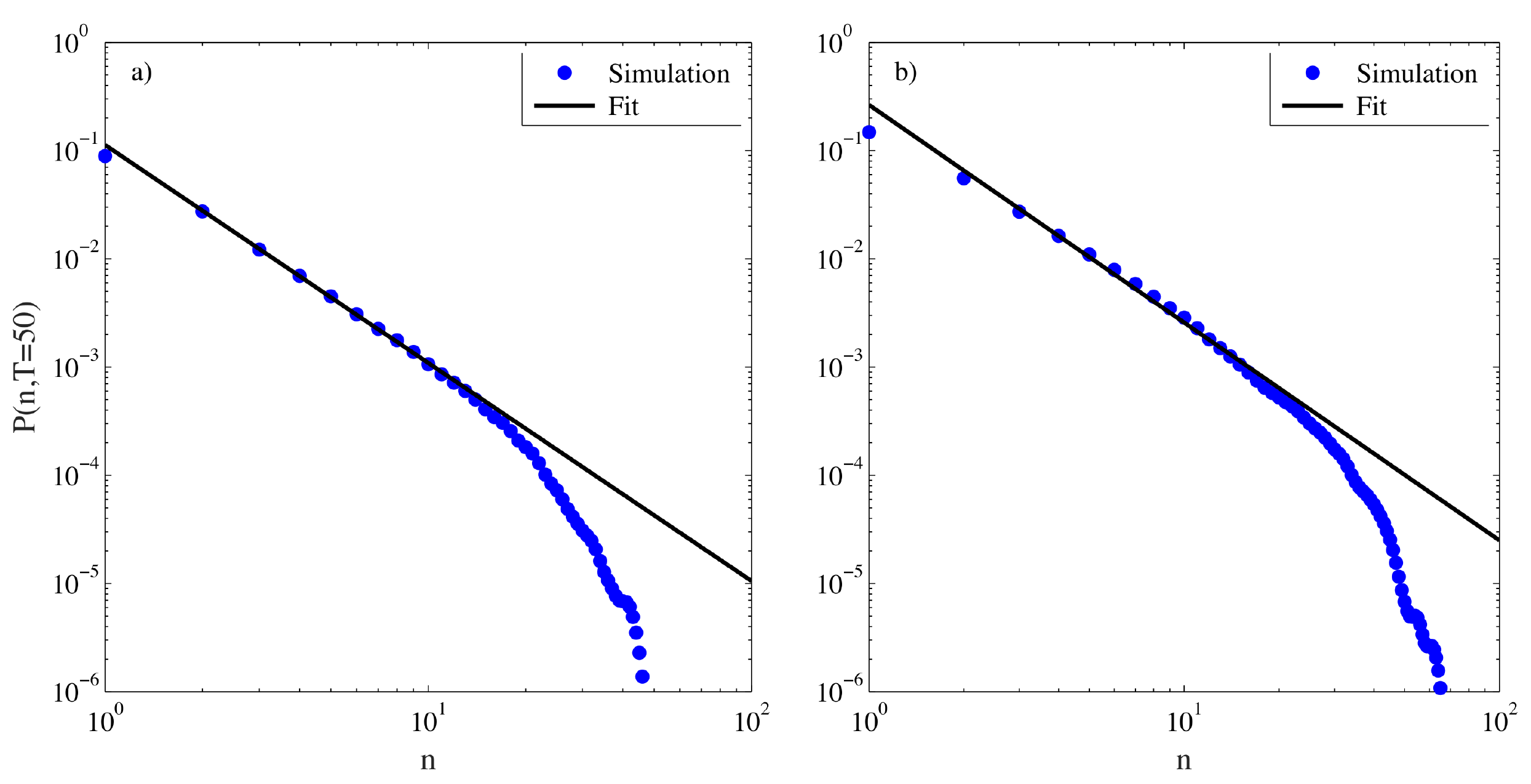}
\caption{Log-log plots of the numerically obtained averaged momentum distributions (blue symbols) via simulations which are close to a real experiment. For $\beta$, $10{,}000$ values taken from a normal distribution with FWHM $\nu=0.06$ are considered. For the $\phi_j$s a uniform distribution is regarded and $k$ is chosen from a limited Cauchy distribution for each repetition. The used limits are $k_{\rm max}=5$ and in $(a)$ $k_{\rm min}=0$ and in $(b)$ $k_{\rm min}=0.1$. $R=1{,}000$ realizations are considered at $T=50$ for $n_0=0$. Both panels also show a power law with exponent 2, corresponding to a Cauchy distribution, fitted to the data (black solid line).}
\label{fig:7}
\end{figure}

Figure \ref{fig:7} shows the results of numerical simulations for which all these realistic experimental parameters are taken into account. For $\beta$, $B=10{,}000$ realizations and a normal distribution with $\nu=0.06$ is considered. As above, for the $\phi_j$s a uniform distribution is assumed and $k$ is chosen from a Cauchy distribution with the above limits for each repetition. 
We consider a small upper limit $k_{\rm max}=5$ with $k_{\rm min}=0$ in panel $(a)$. One observes that the limit for the kicking strength leads to a cut-off in the averaged momentum distribution. This is easily explained with the delta approximation for the square of the Bessel function in Eq.~\eqref{eq:bessel_delta_approx}. Since $k$ is limited, the values for $|R_T|$ are limited and hence the momentum distribution is nearly zero for all $|n| \gtrsim |R_T|_{\rm max}$. The cut-off value in the momentum distribution depends linearly on the maximum limit for $k$, this is in accordance with the mean displacement, which is also linearly dependent on $k$.  We verified numerically that there is indeed a linear correlation between the cut-off value in the momentum distribution and the maximum kicking strength (not shown here, see \cite{weiss2014} for details). In panel $(b)$, the result of exactly the same simulation as done for $(a)$ is shown, with the only difference that here also the minimum limit $k_{\rm min}=0.1$ is taken into account. There are small differences between the resulting distributions. In the second case, the power law sets in slightly later. Furthermore, in the second case the power law region is slightly shifted upwards in the vertical direction. The reason for this effect is that by considering a minimum limit, an interval of the Cauchy distribution (from $0$ up to $0.1$) is cut which has a relatively large probability. Hence, due to the new normalization, the values for $k$ in the range from $0.1$ up to $5$ have a higher probability compared to the first case. In other words, the constant in front of the distribution in Eq. \eqref{eq:limits} depends on the specific limits. Consequently, the momentum values resulting from these values of $k$ are more likely.

\subsubsection{Fluctuations in the kicking strength and optimizing the signal-to-noise ratio}
\label{sec:noise}

Another experimental problem is that $k$ is not completely constant during a sequence because of two independent experimental artifacts. The first one is that not all the atoms sit at the same spot and hence they may feel a slightly different intensity of the kick potential. Additionally, there are time-dependent drifts in the experiment, especially when data is taken over a relatively long period. Considering such fluctuations are of the order $\Delta k = \pm 0.1$, consistent with the lower cutoff chosen in the previous subsection, one does not observe significant differences in the results, see ref. \cite{weiss2014} for details. 

In previous atom-optics experiments weak spontaneous emission was changing the quasimomentum of the atoms and also heating them slightly \cite{WGF2003,DarcyPRE,Wimberger2005PRA,gil2013b}. While this is a nuisance, for instance, for the observation of dynamical tunneling \cite{gil2013b}, it should not be a problem in the present context. Changing randomly quasimomentum effectively leads to a broader distribution of quasimomenta, and it is similar to changing randomly the phase of the kick potential (see discussion in sec. \ref{sec:phase}). Since both of these effects do not alter very much our predictions, spontaneous emission should not hinder the observation of power-law tailed momentum distributions.

The final problem we consider here might be noise in the detection procedure (e.g. due to pixel fluctuation in the CCD camera at low contrast), which makes it difficult to see small signals in the momentum distribution for large $|n|$. Since due to all previous considered limitations, the power law obtained in the momentum distribution only covers a small momentum range and for this range the probability ratio $P(n\!=\!0)_{\rm max}$ to $P(n)_{\rm min}$ is about three orders of magnitude, this might not be a problem. However, a possible way to avoid the noise problem is to choose for the $k$ distribution a power law that decays slower than a Cauchy distribution, i.e. $0 < \alpha < 1$. Hence, the ratio $P(n)_{\rm max}$ to $P(n)_{\rm min}$ will not grow too fast, allowing for a better experimental signal. Together with an appropriate choice of the width parameter $\gamma$, see discussion at the end of subsection \ref{sec:kickS}, this should give the possibility for the experimental implementation of steered random walks with a Bose-Einstein condensate. 

 \section{Conclusions}
\label{sec:6}

We have shown how it is possible to steer the average momentum distribution of kicked ultracold atoms
by applying random sequences of kicks with specific distributions of phases and kick
strengths. 

Our central result is contained in formulae \eqref{eq:Pmean} and \eqref{eq:P_final}, respectively. 
They show that the chosen distribution of kick strengths converts into a momentum distribution with 
the same asymptotic scaling. This allows for future experimental investigations of complex random walks, e.g., power-law
distributed L\'evy walks in momentum space. Random walks are an important part of many biological, 
social, and physical systems \cite{malkiel,econ,levy2015,randomwalk,wiersma2008}. Having a robust scheme for the implementation
of complex classical walks with great stability will enable new insights into many non-deterministic transport processes in nature with large fluctuations. 
The robustness of our prediction with respect to experimental limitations is confirmed by
our detailed analysis of the effects of small detunings from quantum resonance, of 
limits in the kick-strength distribution, and of a finite quasimomentum distribution of the atoms 
(see secs. \ref{sec:4} and \ref{sec:5}).

A random choice of the phase of the kicking lattice from the two values $0$ and $\pi$ mimics 
the coin toss of a quantum walk since it effectively changes the sign of the kick potential, and
hence determines the direction of a single kick or step of the walk. Quantum random walks have
been realized in different setups with single particle control 
\cite{meschede2009,PhysRevLett.104.100503,PhysRevLett.104.153602,Preiss2015}. 
Extending our study to include the necessary additional degree of freedom -- to control the single step of the walk 
simultaneously entangled with the center-of-mass momentum degree of freedom of the atoms -- may
permit one the implementation of stable quantum walks with Bose-Einstein condensates with much more
particles. Moreover, in such improved walks with ultracold atoms, the quantum to classical 
transition \cite{PhysRevLett.99.090601,PhysRevA.87.022334} in the walk behavior may be investigated as partly 
anticipated by our Fig. \ref{fig:2}.
 
\begin{acknowledgments}
We thank Italo Guarneri and Mark Sadgrove for valuable discussions. SW gratefully acknowledges
financial support by the FIL program of the Universit\`a di Parma and by the DFG (Grant No. WI 3426/7). 
\end{acknowledgments}

%


\begin{thebibliography}{55}%
\makeatletter
\providecommand \@ifxundefined [1]{%
 \@ifx{#1\undefined}
}%
\providecommand \@ifnum [1]{%
 \ifnum #1\expandafter \@firstoftwo
 \else \expandafter \@secondoftwo
 \fi
}%
\providecommand \@ifx [1]{%
 \ifx #1\expandafter \@firstoftwo
 \else \expandafter \@secondoftwo
 \fi
}%
\providecommand \natexlab [1]{#1}%
\providecommand \enquote  [1]{``#1''}%
\providecommand \bibnamefont  [1]{#1}%
\providecommand \bibfnamefont [1]{#1}%
\providecommand \citenamefont [1]{#1}%
\providecommand \href@noop [0]{\@secondoftwo}%
\providecommand \href [0]{\begingroup \@sanitize@url \@href}%
\providecommand \@href[1]{\@@startlink{#1}\@@href}%
\providecommand \@@href[1]{\endgroup#1\@@endlink}%
\providecommand \@sanitize@url [0]{\catcode `\\12\catcode `\$12\catcode
  `\&12\catcode `\#12\catcode `\^12\catcode `\_12\catcode `\%12\relax}%
\providecommand \@@startlink[1]{}%
\providecommand \@@endlink[0]{}%
\providecommand \url  [0]{\begingroup\@sanitize@url \@url }%
\providecommand \@url [1]{\endgroup\@href {#1}{\urlprefix }}%
\providecommand \urlprefix  [0]{URL }%
\providecommand \Eprint [0]{\href }%
\providecommand \doibase [0]{http://dx.doi.org/}%
\providecommand \selectlanguage [0]{\@gobble}%
\providecommand \bibinfo  [0]{\@secondoftwo}%
\providecommand \bibfield  [0]{\@secondoftwo}%
\providecommand \translation [1]{[#1]}%
\providecommand \BibitemOpen [0]{}%
\providecommand \bibitemStop [0]{}%
\providecommand \bibitemNoStop [0]{.\EOS\space}%
\providecommand \EOS [0]{\spacefactor3000\relax}%
\providecommand \BibitemShut  [1]{\csname bibitem#1\endcsname}%
\let\auto@bib@innerbib\@empty
\bibitem [{\citenamefont {Weiss}(1994)}]{randomwalk}%
  \BibitemOpen
  \bibfield  {author} {\bibinfo {author} {\bibfnamefont {G.~H.}\ \bibnamefont
  {Weiss}},\ }\href@noop {} {\emph {\bibinfo {title} {Aspects and Applications
  of the Random Walk}}}\ (\bibinfo  {publisher} {North Holland},\ \bibinfo
  {address} {Amsterdam},\ \bibinfo {year} {1994})\BibitemShut {NoStop}%
\bibitem [{\citenamefont {Mantegna}\ and\ \citenamefont
  {Stanley}(2000)}]{econ}%
  \BibitemOpen
  \bibfield  {author} {\bibinfo {author} {\bibfnamefont {R.~N.}\ \bibnamefont
  {Mantegna}}\ and\ \bibinfo {author} {\bibfnamefont {H.~E.}\ \bibnamefont
  {Stanley}},\ }\href@noop {} {\emph {\bibinfo {title} {Introduction to
  econophysics: correlations and complexity in finance}}}\ (\bibinfo
  {publisher} {CUP},\ \bibinfo {address} {Cambridge},\ \bibinfo {year}
  {2000})\BibitemShut {NoStop}%
\bibitem [{\citenamefont {Klenke}(2013)}]{probability_theory}%
  \BibitemOpen
  \bibfield  {author} {\bibinfo {author} {\bibfnamefont {A.}~\bibnamefont
  {Klenke}},\ }\href@noop {} {\emph {\bibinfo {title}
  {{Wahrscheinlichkeitstheorie}}}}\ (\bibinfo  {publisher} {Springer},\
  \bibinfo {address} {Berlin},\ \bibinfo {year} {2013})\BibitemShut {NoStop}%
\bibitem [{\citenamefont {Zaburdaev}\ \emph {et~al.}(2015)\citenamefont
  {Zaburdaev}, \citenamefont {Denisov},\ and\ \citenamefont
  {Klafter}}]{levy2015}%
  \BibitemOpen
  \bibfield  {author} {\bibinfo {author} {\bibfnamefont {V.}~\bibnamefont
  {Zaburdaev}}, \bibinfo {author} {\bibfnamefont {S.}~\bibnamefont {Denisov}},
  \ and\ \bibinfo {author} {\bibfnamefont {J.}~\bibnamefont {Klafter}},\ }\href
  {\doibase 10.1103/RevModPhys.87.483} {\bibfield  {journal} {\bibinfo
  {journal} {Rev. Mod. Phys.}\ }\textbf {\bibinfo {volume} {87}},\ \bibinfo
  {pages} {483} (\bibinfo {year} {2015})}\BibitemShut {NoStop}%
\bibitem [{\citenamefont {Barthelemy}\ \emph {et~al.}(2008)\citenamefont
  {Barthelemy}, \citenamefont {Bertolotti},\ and\ \citenamefont
  {Wiersma}}]{wiersma2008}%
  \BibitemOpen
  \bibfield  {author} {\bibinfo {author} {\bibfnamefont {P.}~\bibnamefont
  {Barthelemy}}, \bibinfo {author} {\bibfnamefont {J.}~\bibnamefont
  {Bertolotti}}, \ and\ \bibinfo {author} {\bibfnamefont {D.~S.}\ \bibnamefont
  {Wiersma}},\ }\href@noop {} {\bibfield  {journal} {\bibinfo  {journal}
  {Nature (London)}\ }\textbf {\bibinfo {volume} {453}},\ \bibinfo {pages}
  {495} (\bibinfo {year} {2008})}\BibitemShut {NoStop}%
\bibitem [{\citenamefont {Buonsante}\ \emph {et~al.}(2011)\citenamefont
  {Buonsante}, \citenamefont {Burioni},\ and\ \citenamefont {Vezzani}}]{pr2}%
  \BibitemOpen
  \bibfield  {author} {\bibinfo {author} {\bibfnamefont {P.}~\bibnamefont
  {Buonsante}}, \bibinfo {author} {\bibfnamefont {R.}~\bibnamefont {Burioni}},
  \ and\ \bibinfo {author} {\bibfnamefont {A.}~\bibnamefont {Vezzani}},\ }\href
  {\doibase 10.1103/PhysRevE.84.021105} {\bibfield  {journal} {\bibinfo
  {journal} {Phys. Rev. E}\ }\textbf {\bibinfo {volume} {84}},\ \bibinfo
  {pages} {021105} (\bibinfo {year} {2011})}\BibitemShut {NoStop}%
\bibitem [{\citenamefont {Burresi}\ \emph {et~al.}(2012)\citenamefont
  {Burresi}, \citenamefont {Radhalakshmi}, \citenamefont {Savo}, \citenamefont
  {Bertolotti}, \citenamefont {Vynck},\ and\ \citenamefont
  {Wiersma}}]{wiersma2012}%
  \BibitemOpen
  \bibfield  {author} {\bibinfo {author} {\bibfnamefont {M.}~\bibnamefont
  {Burresi}}, \bibinfo {author} {\bibfnamefont {V.}~\bibnamefont
  {Radhalakshmi}}, \bibinfo {author} {\bibfnamefont {R.}~\bibnamefont {Savo}},
  \bibinfo {author} {\bibfnamefont {J.}~\bibnamefont {Bertolotti}}, \bibinfo
  {author} {\bibfnamefont {K.}~\bibnamefont {Vynck}}, \ and\ \bibinfo {author}
  {\bibfnamefont {D.~S.}\ \bibnamefont {Wiersma}},\ }\href {\doibase
  10.1103/PhysRevLett.108.110604} {\bibfield  {journal} {\bibinfo  {journal}
  {Phys. Rev. Lett.}\ }\textbf {\bibinfo {volume} {108}},\ \bibinfo {pages}
  {110604} (\bibinfo {year} {2012})}\BibitemShut {NoStop}%
\bibitem [{\citenamefont {Svensson}\ \emph {et~al.}(2013)\citenamefont
  {Svensson}, \citenamefont {Vynck}, \citenamefont {Grisi}, \citenamefont
  {Savo}, \citenamefont {Burresi},\ and\ \citenamefont
  {Wiersma}}]{wiersma2013}%
  \BibitemOpen
  \bibfield  {author} {\bibinfo {author} {\bibfnamefont {T.}~\bibnamefont
  {Svensson}}, \bibinfo {author} {\bibfnamefont {K.}~\bibnamefont {Vynck}},
  \bibinfo {author} {\bibfnamefont {M.}~\bibnamefont {Grisi}}, \bibinfo
  {author} {\bibfnamefont {R.}~\bibnamefont {Savo}}, \bibinfo {author}
  {\bibfnamefont {M.}~\bibnamefont {Burresi}}, \ and\ \bibinfo {author}
  {\bibfnamefont {D.~S.}\ \bibnamefont {Wiersma}},\ }\href {\doibase
  10.1103/PhysRevE.87.022120} {\bibfield  {journal} {\bibinfo  {journal} {Phys.
  Rev. E}\ }\textbf {\bibinfo {volume} {87}},\ \bibinfo {pages} {022120}
  (\bibinfo {year} {2013})}\BibitemShut {NoStop}%
\bibitem [{\citenamefont {Burioni}\ \emph {et~al.}(2014)\citenamefont
  {Burioni}, \citenamefont {Ubaldi},\ and\ \citenamefont {Vezzani}}]{pr1}%
  \BibitemOpen
  \bibfield  {author} {\bibinfo {author} {\bibfnamefont {R.}~\bibnamefont
  {Burioni}}, \bibinfo {author} {\bibfnamefont {E.}~\bibnamefont {Ubaldi}}, \
  and\ \bibinfo {author} {\bibfnamefont {A.}~\bibnamefont {Vezzani}},\ }\href
  {\doibase 10.1103/PhysRevE.89.022135} {\bibfield  {journal} {\bibinfo
  {journal} {Phys. Rev. E}\ }\textbf {\bibinfo {volume} {89}},\ \bibinfo
  {pages} {022135} (\bibinfo {year} {2014})}\BibitemShut {NoStop}%
\bibitem [{\citenamefont {Malkiel}(2012)}]{malkiel}%
  \BibitemOpen
  \bibfield  {author} {\bibinfo {author} {\bibfnamefont {B.}~\bibnamefont
  {Malkiel}},\ }\href@noop {} {\emph {\bibinfo {title} {A Random Walk Down Wall
  Street}}}\ (\bibinfo  {publisher} {Norton},\ \bibinfo {address} {New York},\
  \bibinfo {year} {2012})\BibitemShut {NoStop}%
\bibitem [{\citenamefont {Aharonov}\ \emph {et~al.}(1993)\citenamefont
  {Aharonov}, \citenamefont {Davidovich},\ and\ \citenamefont
  {Zagury}}]{PhysRevA.48.1687}%
  \BibitemOpen
  \bibfield  {author} {\bibinfo {author} {\bibfnamefont {Y.}~\bibnamefont
  {Aharonov}}, \bibinfo {author} {\bibfnamefont {L.}~\bibnamefont
  {Davidovich}}, \ and\ \bibinfo {author} {\bibfnamefont {N.}~\bibnamefont
  {Zagury}},\ }\href {\doibase 10.1103/PhysRevA.48.1687} {\bibfield  {journal}
  {\bibinfo  {journal} {Phys. Rev. A}\ }\textbf {\bibinfo {volume} {48}},\
  \bibinfo {pages} {1687} (\bibinfo {year} {1993})}\BibitemShut {NoStop}%
\bibitem [{\citenamefont {Kempe}(2003)}]{kempe2003}%
  \BibitemOpen
  \bibfield  {author} {\bibinfo {author} {\bibfnamefont {J.}~\bibnamefont
  {Kempe}},\ }\href@noop {} {\bibfield  {journal} {\bibinfo  {journal}
  {Contemporary Physics}\ }\textbf {\bibinfo {volume} {44}},\ \bibinfo {pages}
  {307} (\bibinfo {year} {2003})}\BibitemShut {NoStop}%
\bibitem [{\citenamefont {Marksteiner}\ \emph {et~al.}(1996)\citenamefont
  {Marksteiner}, \citenamefont {Ellinger},\ and\ \citenamefont
  {Zoller}}]{zoller1996}%
  \BibitemOpen
  \bibfield  {author} {\bibinfo {author} {\bibfnamefont {S.}~\bibnamefont
  {Marksteiner}}, \bibinfo {author} {\bibfnamefont {K.}~\bibnamefont
  {Ellinger}}, \ and\ \bibinfo {author} {\bibfnamefont {P.}~\bibnamefont
  {Zoller}},\ }\href {\doibase 10.1103/PhysRevA.53.3409} {\bibfield  {journal}
  {\bibinfo  {journal} {Phys. Rev. A}\ }\textbf {\bibinfo {volume} {53}},\
  \bibinfo {pages} {3409} (\bibinfo {year} {1996})}\BibitemShut {NoStop}%
\bibitem [{\citenamefont {Micciche}\ \emph {et~al.}(2013)\citenamefont
  {Micciche}, \citenamefont {Buchleitner}, \citenamefont {Lillo}, \citenamefont
  {Mantegna}, \citenamefont {Paul},\ and\ \citenamefont {Wimberger}}]{pa2013}%
  \BibitemOpen
  \bibfield  {author} {\bibinfo {author} {\bibfnamefont {S.}~\bibnamefont
  {Micciche}}, \bibinfo {author} {\bibfnamefont {A.}~\bibnamefont
  {Buchleitner}}, \bibinfo {author} {\bibfnamefont {F.}~\bibnamefont {Lillo}},
  \bibinfo {author} {\bibfnamefont {R.~N.}\ \bibnamefont {Mantegna}}, \bibinfo
  {author} {\bibfnamefont {T.}~\bibnamefont {Paul}}, \ and\ \bibinfo {author}
  {\bibfnamefont {S.}~\bibnamefont {Wimberger}},\ }\href
  {http://stacks.iop.org/1367-2630/15/i=3/a=033033} {\bibfield  {journal}
  {\bibinfo  {journal} {New Journal of Physics}\ }\textbf {\bibinfo {volume}
  {15}},\ \bibinfo {pages} {033033} (\bibinfo {year} {2013})}\BibitemShut
  {NoStop}%
\bibitem [{\citenamefont {Ma}\ \emph {et~al.}(2006)\citenamefont {Ma},
  \citenamefont {Burnett}, \citenamefont {d'Arcy},\ and\ \citenamefont
  {Gardiner}}]{PhysRevA.73.013401}%
  \BibitemOpen
  \bibfield  {author} {\bibinfo {author} {\bibfnamefont {Z.-Y.}\ \bibnamefont
  {Ma}}, \bibinfo {author} {\bibfnamefont {K.}~\bibnamefont {Burnett}},
  \bibinfo {author} {\bibfnamefont {M.~B.}\ \bibnamefont {d'Arcy}}, \ and\
  \bibinfo {author} {\bibfnamefont {S.~A.}\ \bibnamefont {Gardiner}},\ }\href
  {\doibase 10.1103/PhysRevA.73.013401} {\bibfield  {journal} {\bibinfo
  {journal} {Phys. Rev. A}\ }\textbf {\bibinfo {volume} {73}},\ \bibinfo
  {pages} {013401} (\bibinfo {year} {2006})}\BibitemShut {NoStop}%
\bibitem [{\citenamefont {Schomerus}\ and\ \citenamefont
  {Lutz}(2008)}]{lutz2007}%
  \BibitemOpen
  \bibfield  {author} {\bibinfo {author} {\bibfnamefont {H.}~\bibnamefont
  {Schomerus}}\ and\ \bibinfo {author} {\bibfnamefont {E.}~\bibnamefont
  {Lutz}},\ }\href {\doibase 10.1103/PhysRevA.77.062113} {\bibfield  {journal}
  {\bibinfo  {journal} {Phys. Rev. A}\ }\textbf {\bibinfo {volume} {77}},\
  \bibinfo {pages} {062113} (\bibinfo {year} {2008})}\BibitemShut {NoStop}%
\bibitem [{\citenamefont {Schreiber}\ \emph {et~al.}(2010)\citenamefont
  {Schreiber}, \citenamefont {Cassemiro}, \citenamefont
  {Poto\ifmmode~\check{c}\else \v{c}\fi{}ek}, \citenamefont {G\'abris},
  \citenamefont {Mosley}, \citenamefont {Andersson}, \citenamefont {Jex},\ and\
  \citenamefont {Silberhorn}}]{silberhorn1}%
  \BibitemOpen
  \bibfield  {author} {\bibinfo {author} {\bibfnamefont {A.}~\bibnamefont
  {Schreiber}}, \bibinfo {author} {\bibfnamefont {K.~N.}\ \bibnamefont
  {Cassemiro}}, \bibinfo {author} {\bibfnamefont {V.}~\bibnamefont
  {Poto\ifmmode~\check{c}\else \v{c}\fi{}ek}}, \bibinfo {author} {\bibfnamefont
  {A.}~\bibnamefont {G\'abris}}, \bibinfo {author} {\bibfnamefont {P.~J.}\
  \bibnamefont {Mosley}}, \bibinfo {author} {\bibfnamefont {E.}~\bibnamefont
  {Andersson}}, \bibinfo {author} {\bibfnamefont {I.}~\bibnamefont {Jex}}, \
  and\ \bibinfo {author} {\bibfnamefont {C.}~\bibnamefont {Silberhorn}},\
  }\href {\doibase 10.1103/PhysRevLett.104.050502} {\bibfield  {journal}
  {\bibinfo  {journal} {Phys. Rev. Lett.}\ }\textbf {\bibinfo {volume} {104}},\
  \bibinfo {pages} {050502} (\bibinfo {year} {2010})}\BibitemShut {NoStop}%
\bibitem [{\citenamefont {Schreiber}\ \emph {et~al.}(2011)\citenamefont
  {Schreiber}, \citenamefont {Cassemiro}, \citenamefont
  {Poto\ifmmode~\check{c}\else \v{c}\fi{}ek}, \citenamefont {G\'abris},
  \citenamefont {Jex},\ and\ \citenamefont {Silberhorn}}]{silberhorn2}%
  \BibitemOpen
  \bibfield  {author} {\bibinfo {author} {\bibfnamefont {A.}~\bibnamefont
  {Schreiber}}, \bibinfo {author} {\bibfnamefont {K.~N.}\ \bibnamefont
  {Cassemiro}}, \bibinfo {author} {\bibfnamefont {V.}~\bibnamefont
  {Poto\ifmmode~\check{c}\else \v{c}\fi{}ek}}, \bibinfo {author} {\bibfnamefont
  {A.}~\bibnamefont {G\'abris}}, \bibinfo {author} {\bibfnamefont
  {I.}~\bibnamefont {Jex}}, \ and\ \bibinfo {author} {\bibfnamefont
  {C.}~\bibnamefont {Silberhorn}},\ }\href {\doibase
  10.1103/PhysRevLett.106.180403} {\bibfield  {journal} {\bibinfo  {journal}
  {Phys. Rev. Lett.}\ }\textbf {\bibinfo {volume} {106}},\ \bibinfo {pages}
  {180403} (\bibinfo {year} {2011})}\BibitemShut {NoStop}%
\bibitem [{\citenamefont {Schlunk}\ \emph {et~al.}(2003)\citenamefont
  {Schlunk}, \citenamefont {d'Arcy}, \citenamefont {Gardiner}, \citenamefont
  {Cassettari}, \citenamefont {Godun},\ and\ \citenamefont {Summy}}]{schlunk1}%
  \BibitemOpen
  \bibfield  {author} {\bibinfo {author} {\bibfnamefont {S.}~\bibnamefont
  {Schlunk}}, \bibinfo {author} {\bibfnamefont {M.~B.}\ \bibnamefont {d'Arcy}},
  \bibinfo {author} {\bibfnamefont {S.~A.}\ \bibnamefont {Gardiner}}, \bibinfo
  {author} {\bibfnamefont {D.}~\bibnamefont {Cassettari}}, \bibinfo {author}
  {\bibfnamefont {R.~M.}\ \bibnamefont {Godun}}, \ and\ \bibinfo {author}
  {\bibfnamefont {G.~S.}\ \bibnamefont {Summy}},\ }\href {\doibase
  10.1103/PhysRevLett.90.054101} {\bibfield  {journal} {\bibinfo  {journal}
  {Phys. Rev. Lett.}\ }\textbf {\bibinfo {volume} {90}},\ \bibinfo {pages}
  {054101} (\bibinfo {year} {2003})}\BibitemShut {NoStop}%
\bibitem [{\citenamefont {Wimberger}\ and\ \citenamefont
  {Buchleitner}(2006)}]{WB2006}%
  \BibitemOpen
  \bibfield  {author} {\bibinfo {author} {\bibfnamefont {S.}~\bibnamefont
  {Wimberger}}\ and\ \bibinfo {author} {\bibfnamefont {A.}~\bibnamefont
  {Buchleitner}},\ }\href {http://stacks.iop.org/0953-4075/39/i=7/a=L01}
  {\bibfield  {journal} {\bibinfo  {journal} {Journal of Physics B: Atomic,
  Molecular and Optical Physics}\ }\textbf {\bibinfo {volume} {39}},\ \bibinfo
  {pages} {L145} (\bibinfo {year} {2006})}\BibitemShut {NoStop}%
\bibitem [{\citenamefont {Sadgrove}\ \emph {et~al.}(2012)\citenamefont
  {Sadgrove}, \citenamefont {Wimberger},\ and\ \citenamefont
  {Nakagawa}}]{sadgrove2012}%
  \BibitemOpen
  \bibfield  {author} {\bibinfo {author} {\bibfnamefont {M.}~\bibnamefont
  {Sadgrove}}, \bibinfo {author} {\bibfnamefont {S.}~\bibnamefont {Wimberger}},
  \ and\ \bibinfo {author} {\bibfnamefont {K.}~\bibnamefont {Nakagawa}},\
  }\href {http://dx.doi.org/10.1140/epjd/e2012-20578-6} {\bibfield  {journal}
  {\bibinfo  {journal} {The European Physical Journal D}\ }\textbf {\bibinfo
  {volume} {66}},\ \bibinfo {eid} {155} (\bibinfo {year} {2012})}\BibitemShut
  {NoStop}%
\bibitem [{\citenamefont {Raizen}(1999)}]{RaizenAdv}%
  \BibitemOpen
  \bibfield  {author} {\bibinfo {author} {\bibfnamefont {M.~G.}\ \bibnamefont
  {Raizen}},\ }\href@noop {} {\bibfield  {journal} {\bibinfo  {journal} {Adv.
  At. Mol. Phys.}\ }\textbf {\bibinfo {volume} {41}},\ \bibinfo {pages} {43}
  (\bibinfo {year} {1999})}\BibitemShut {NoStop}%
\bibitem [{\citenamefont {Sadgrove}\ and\ \citenamefont
  {Wimberger}(2011)}]{SW2011}%
  \BibitemOpen
  \bibfield  {author} {\bibinfo {author} {\bibfnamefont {M.}~\bibnamefont
  {Sadgrove}}\ and\ \bibinfo {author} {\bibfnamefont {S.}~\bibnamefont
  {Wimberger}},\ }\href@noop {} {\bibfield  {journal} {\bibinfo  {journal}
  {Adv. At. Mol. Opt. Phys.}\ }\textbf {\bibinfo {volume} {60}},\ \bibinfo
  {pages} {315} (\bibinfo {year} {2011})}\BibitemShut {NoStop}%
\bibitem [{\citenamefont {Izrailev}\ and\ \citenamefont
  {Shepelyansky}(1979)}]{shepelyansky1}%
  \BibitemOpen
  \bibfield  {author} {\bibinfo {author} {\bibfnamefont {F.~M.}\ \bibnamefont
  {Izrailev}}\ and\ \bibinfo {author} {\bibfnamefont {D.~L.}\ \bibnamefont
  {Shepelyansky}},\ }\href@noop {} {\bibfield  {journal} {\bibinfo  {journal}
  {Sov. Phys. Dokl.}\ }\textbf {\bibinfo {volume} {24}},\ \bibinfo {pages}
  {996} (\bibinfo {year} {1979})}\BibitemShut {NoStop}%
\bibitem [{\citenamefont {Izrailev}(1990)}]{Izr1990}%
  \BibitemOpen
  \bibfield  {author} {\bibinfo {author} {\bibfnamefont {F.~M.}\ \bibnamefont
  {Izrailev}},\ }\href {\doibase DOI: 10.1016/0370-1573(90)90067-C} {\bibfield
  {journal} {\bibinfo  {journal} {Physics Reports}\ }\textbf {\bibinfo {volume}
  {196}},\ \bibinfo {pages} {299 } (\bibinfo {year} {1990})}\BibitemShut
  {NoStop}%
\bibitem [{\citenamefont {Wimberger}\ \emph {et~al.}(2003)\citenamefont
  {Wimberger}, \citenamefont {Guarneri},\ and\ \citenamefont
  {Fishman}}]{WGF2003}%
  \BibitemOpen
  \bibfield  {author} {\bibinfo {author} {\bibfnamefont {S.}~\bibnamefont
  {Wimberger}}, \bibinfo {author} {\bibfnamefont {I.}~\bibnamefont {Guarneri}},
  \ and\ \bibinfo {author} {\bibfnamefont {S.}~\bibnamefont {Fishman}},\ }\href
  {http://stacks.iop.org/0951-7715/16/i=4/a=312} {\bibfield  {journal}
  {\bibinfo  {journal} {Nonlinearity}\ }\textbf {\bibinfo {volume} {16}},\
  \bibinfo {pages} {1381} (\bibinfo {year} {2003})}\BibitemShut {NoStop}%
\bibitem [{\citenamefont {Fishman}\ \emph {et~al.}(2003)\citenamefont
  {Fishman}, \citenamefont {Guarneri},\ and\ \citenamefont
  {Rebuzzini}}]{FGR2003}%
  \BibitemOpen
  \bibfield  {author} {\bibinfo {author} {\bibfnamefont {S.}~\bibnamefont
  {Fishman}}, \bibinfo {author} {\bibfnamefont {I.}~\bibnamefont {Guarneri}}, \
  and\ \bibinfo {author} {\bibfnamefont {L.}~\bibnamefont {Rebuzzini}},\
  }\href@noop {} {\bibfield  {journal} {\bibinfo  {journal} {Journal of
  Statistical Physics}\ }\textbf {\bibinfo {volume} {110}},\ \bibinfo {pages}
  {911} (\bibinfo {year} {2003})}\BibitemShut {NoStop}%
\bibitem [{\citenamefont {Guarneri}\ \emph {et~al.}(2006)\citenamefont
  {Guarneri}, \citenamefont {Rebuzzini},\ and\ \citenamefont
  {Fishman}}]{FGR2006}%
  \BibitemOpen
  \bibfield  {author} {\bibinfo {author} {\bibfnamefont {I.}~\bibnamefont
  {Guarneri}}, \bibinfo {author} {\bibfnamefont {L.}~\bibnamefont {Rebuzzini}},
  \ and\ \bibinfo {author} {\bibfnamefont {S.}~\bibnamefont {Fishman}},\ }\href
  {http://stacks.iop.org/0951-7715/19/i=5/a=006} {\bibfield  {journal}
  {\bibinfo  {journal} {Nonlinearity}\ }\textbf {\bibinfo {volume} {19}},\
  \bibinfo {pages} {1141} (\bibinfo {year} {2006})}\BibitemShut {NoStop}%
\bibitem [{\citenamefont {Lundh}\ and\ \citenamefont
  {Wallin}(2005)}]{PhysRevLett.94.110603}%
  \BibitemOpen
  \bibfield  {author} {\bibinfo {author} {\bibfnamefont {E.}~\bibnamefont
  {Lundh}}\ and\ \bibinfo {author} {\bibfnamefont {M.}~\bibnamefont {Wallin}},\
  }\href {\doibase 10.1103/PhysRevLett.94.110603} {\bibfield  {journal}
  {\bibinfo  {journal} {Phys. Rev. Lett.}\ }\textbf {\bibinfo {volume} {94}},\
  \bibinfo {pages} {110603} (\bibinfo {year} {2005})}\BibitemShut {NoStop}%
\bibitem [{\citenamefont {Sadgrove}\ \emph {et~al.}(2007)\citenamefont
  {Sadgrove}, \citenamefont {Horikoshi}, \citenamefont {Sekimura},\ and\
  \citenamefont {Nakagawa}}]{sadgrove2007}%
  \BibitemOpen
  \bibfield  {author} {\bibinfo {author} {\bibfnamefont {M.}~\bibnamefont
  {Sadgrove}}, \bibinfo {author} {\bibfnamefont {M.}~\bibnamefont {Horikoshi}},
  \bibinfo {author} {\bibfnamefont {T.}~\bibnamefont {Sekimura}}, \ and\
  \bibinfo {author} {\bibfnamefont {K.}~\bibnamefont {Nakagawa}},\ }\href
  {\doibase 10.1103/PhysRevLett.99.043002} {\bibfield  {journal} {\bibinfo
  {journal} {Phys. Rev. Lett.}\ }\textbf {\bibinfo {volume} {99}},\ \bibinfo
  {pages} {043002} (\bibinfo {year} {2007})}\BibitemShut {NoStop}%
\bibitem [{\citenamefont {Sadgrove}\ and\ \citenamefont
  {Wimberger}(2009)}]{SadgroveWimberger2009}%
  \BibitemOpen
  \bibfield  {author} {\bibinfo {author} {\bibfnamefont {M.}~\bibnamefont
  {Sadgrove}}\ and\ \bibinfo {author} {\bibfnamefont {S.}~\bibnamefont
  {Wimberger}},\ }\href {http://stacks.iop.org/1367-2630/11/i=8/a=083027}
  {\bibfield  {journal} {\bibinfo  {journal} {New Journal of Physics}\ }\textbf
  {\bibinfo {volume} {11}},\ \bibinfo {pages} {083027} (\bibinfo {year}
  {2009})}\BibitemShut {NoStop}%
\bibitem [{\citenamefont {Sadgrove}\ \emph {et~al.}(2013)\citenamefont
  {Sadgrove}, \citenamefont {Schell}, \citenamefont {Nakagawa},\ and\
  \citenamefont {Wimberger}}]{sad2013}%
  \BibitemOpen
  \bibfield  {author} {\bibinfo {author} {\bibfnamefont {M.}~\bibnamefont
  {Sadgrove}}, \bibinfo {author} {\bibfnamefont {T.}~\bibnamefont {Schell}},
  \bibinfo {author} {\bibfnamefont {K.}~\bibnamefont {Nakagawa}}, \ and\
  \bibinfo {author} {\bibfnamefont {S.}~\bibnamefont {Wimberger}},\ }\href
  {\doibase 10.1103/PhysRevA.87.013631} {\bibfield  {journal} {\bibinfo
  {journal} {Phys. Rev. A}\ }\textbf {\bibinfo {volume} {87}},\ \bibinfo
  {pages} {013631} (\bibinfo {year} {2013})}\BibitemShut {NoStop}%
\bibitem [{\citenamefont {Dana}\ \emph {et~al.}(2008)\citenamefont {Dana},
  \citenamefont {Ramareddy}, \citenamefont {Talukdar},\ and\ \citenamefont
  {Summy}}]{gil2008}%
  \BibitemOpen
  \bibfield  {author} {\bibinfo {author} {\bibfnamefont {I.}~\bibnamefont
  {Dana}}, \bibinfo {author} {\bibfnamefont {V.}~\bibnamefont {Ramareddy}},
  \bibinfo {author} {\bibfnamefont {I.}~\bibnamefont {Talukdar}}, \ and\
  \bibinfo {author} {\bibfnamefont {G.~S.}\ \bibnamefont {Summy}},\ }\href
  {\doibase 10.1103/PhysRevLett.100.024103} {\bibfield  {journal} {\bibinfo
  {journal} {Phys. Rev. Lett.}\ }\textbf {\bibinfo {volume} {100}},\ \bibinfo
  {pages} {024103} (\bibinfo {year} {2008})}\BibitemShut {NoStop}%
\bibitem [{\citenamefont {d'Arcy}\ \emph {et~al.}(2001)\citenamefont {d'Arcy},
  \citenamefont {Godun}, \citenamefont {Oberthaler}, \citenamefont
  {Cassettari},\ and\ \citenamefont {Summy}}]{darcy2001}%
  \BibitemOpen
  \bibfield  {author} {\bibinfo {author} {\bibfnamefont {M.~B.}\ \bibnamefont
  {d'Arcy}}, \bibinfo {author} {\bibfnamefont {R.~M.}\ \bibnamefont {Godun}},
  \bibinfo {author} {\bibfnamefont {M.~K.}\ \bibnamefont {Oberthaler}},
  \bibinfo {author} {\bibfnamefont {D.}~\bibnamefont {Cassettari}}, \ and\
  \bibinfo {author} {\bibfnamefont {G.~S.}\ \bibnamefont {Summy}},\ }\href
  {\doibase 10.1103/PhysRevLett.87.074102} {\bibfield  {journal} {\bibinfo
  {journal} {Phys. Rev. Lett.}\ }\textbf {\bibinfo {volume} {87}},\ \bibinfo
  {pages} {074102} (\bibinfo {year} {2001})}\BibitemShut {NoStop}%
\bibitem [{\citenamefont {Sadgrove}\ \emph {et~al.}(2004)\citenamefont
  {Sadgrove}, \citenamefont {Hilliard}, \citenamefont {Mullins}, \citenamefont
  {Parkins},\ and\ \citenamefont {Leonhardt}}]{SadgrovePRE}%
  \BibitemOpen
  \bibfield  {author} {\bibinfo {author} {\bibfnamefont {M.}~\bibnamefont
  {Sadgrove}}, \bibinfo {author} {\bibfnamefont {A.}~\bibnamefont {Hilliard}},
  \bibinfo {author} {\bibfnamefont {T.}~\bibnamefont {Mullins}}, \bibinfo
  {author} {\bibfnamefont {S.}~\bibnamefont {Parkins}}, \ and\ \bibinfo
  {author} {\bibfnamefont {R.}~\bibnamefont {Leonhardt}},\ }\href {\doibase
  10.1103/PhysRevE.70.036217} {\bibfield  {journal} {\bibinfo  {journal} {Phys.
  Rev. E}\ }\textbf {\bibinfo {volume} {70}},\ \bibinfo {pages} {036217}
  (\bibinfo {year} {2004})}\BibitemShut {NoStop}%
\bibitem [{\citenamefont {Sadgrove}\ \emph {et~al.}(2008)\citenamefont
  {Sadgrove}, \citenamefont {Wimberger}, \citenamefont {Parkins},\ and\
  \citenamefont {Leonhardt}}]{sadgrove2008}%
  \BibitemOpen
  \bibfield  {author} {\bibinfo {author} {\bibfnamefont {M.}~\bibnamefont
  {Sadgrove}}, \bibinfo {author} {\bibfnamefont {S.}~\bibnamefont {Wimberger}},
  \bibinfo {author} {\bibfnamefont {S.}~\bibnamefont {Parkins}}, \ and\
  \bibinfo {author} {\bibfnamefont {R.}~\bibnamefont {Leonhardt}},\ }\href
  {\doibase 10.1103/PhysRevE.78.025206} {\bibfield  {journal} {\bibinfo
  {journal} {Phys. Rev. E}\ }\textbf {\bibinfo {volume} {78}},\ \bibinfo
  {pages} {025206} (\bibinfo {year} {2008})}\BibitemShut {NoStop}%
\bibitem [{\citenamefont {Romanelli}\ and\ \citenamefont
  {Hernandez}(2010)}]{romanelli2010}%
  \BibitemOpen
  \bibfield  {author} {\bibinfo {author} {\bibfnamefont {A.}~\bibnamefont
  {Romanelli}}\ and\ \bibinfo {author} {\bibfnamefont {G.}~\bibnamefont
  {Hernandez}},\ }\href@noop {} {\bibfield  {journal} {\bibinfo  {journal}
  {Physica A}\ }\textbf {\bibinfo {volume} {389}},\ \bibinfo {pages} {3420}
  (\bibinfo {year} {2010})}\BibitemShut {NoStop}%
\bibitem [{\citenamefont {Abramowitz}\ and\ \citenamefont
  {Stegun}(1972)}]{AS72}%
  \BibitemOpen
  \bibfield  {author} {\bibinfo {author} {\bibfnamefont {M.}~\bibnamefont
  {Abramowitz}}\ and\ \bibinfo {author} {\bibfnamefont {I.~A.}\ \bibnamefont
  {Stegun}},\ }\href@noop {} {\emph {\bibinfo {title} {Handbook of mathematical
  functions}}}\ (\bibinfo  {publisher} {Dover},\ \bibinfo {address} {New
  York},\ \bibinfo {year} {1972})\BibitemShut {NoStop}%
\bibitem [{\citenamefont {Wei\ss}(2014)}]{weiss2014}%
  \BibitemOpen
  \bibfield  {author} {\bibinfo {author} {\bibfnamefont {M.}~\bibnamefont
  {Wei\ss}},\ }\emph {\bibinfo {title} {Steering random walks through quantum
  resonance}},\ \href@noop {} {\bibinfo {type} {{B.S. Thesis}}},\ \bibinfo
  {school} {University of Heidelberg} (\bibinfo {year} {2014})\BibitemShut
  {NoStop}%
\bibitem [{\citenamefont {Talukdar}\ \emph {et~al.}(2010)\citenamefont
  {Talukdar}, \citenamefont {Shrestha},\ and\ \citenamefont {Summy}}]{gil2010}%
  \BibitemOpen
  \bibfield  {author} {\bibinfo {author} {\bibfnamefont {I.}~\bibnamefont
  {Talukdar}}, \bibinfo {author} {\bibfnamefont {R.}~\bibnamefont {Shrestha}},
  \ and\ \bibinfo {author} {\bibfnamefont {G.~S.}\ \bibnamefont {Summy}},\
  }\href {\doibase 10.1103/PhysRevLett.105.054103} {\bibfield  {journal}
  {\bibinfo  {journal} {Phys. Rev. Lett.}\ }\textbf {\bibinfo {volume} {105}},\
  \bibinfo {pages} {054103} (\bibinfo {year} {2010})}\BibitemShut {NoStop}%
\bibitem [{\citenamefont {Shrestha}\ \emph
  {et~al.}(2013{\natexlab{a}})\citenamefont {Shrestha}, \citenamefont
  {Wimberger}, \citenamefont {Ni}, \citenamefont {Lam},\ and\ \citenamefont
  {Summy}}]{gil2013a}%
  \BibitemOpen
  \bibfield  {author} {\bibinfo {author} {\bibfnamefont {R.~K.}\ \bibnamefont
  {Shrestha}}, \bibinfo {author} {\bibfnamefont {S.}~\bibnamefont {Wimberger}},
  \bibinfo {author} {\bibfnamefont {J.}~\bibnamefont {Ni}}, \bibinfo {author}
  {\bibfnamefont {W.~K.}\ \bibnamefont {Lam}}, \ and\ \bibinfo {author}
  {\bibfnamefont {G.~S.}\ \bibnamefont {Summy}},\ }\href {\doibase
  10.1103/PhysRevE.87.020902} {\bibfield  {journal} {\bibinfo  {journal} {Phys.
  Rev. E}\ }\textbf {\bibinfo {volume} {87}},\ \bibinfo {pages} {020902}
  (\bibinfo {year} {2013}{\natexlab{a}})}\BibitemShut {NoStop}%
\bibitem [{\citenamefont {White}\ \emph {et~al.}(2013)\citenamefont {White},
  \citenamefont {Ruddell},\ and\ \citenamefont {Hoogerland}}]{hoogerland2013}%
  \BibitemOpen
  \bibfield  {author} {\bibinfo {author} {\bibfnamefont {D.~H.}\ \bibnamefont
  {White}}, \bibinfo {author} {\bibfnamefont {S.~K.}\ \bibnamefont {Ruddell}},
  \ and\ \bibinfo {author} {\bibfnamefont {M.~D.}\ \bibnamefont {Hoogerland}},\
  }\href {\doibase 10.1103/PhysRevA.88.063603} {\bibfield  {journal} {\bibinfo
  {journal} {Phys. Rev. A}\ }\textbf {\bibinfo {volume} {88}},\ \bibinfo
  {pages} {063603} (\bibinfo {year} {2013})}\BibitemShut {NoStop}%
\bibitem [{\citenamefont {Chirikov}(1979)}]{chirikov}%
  \BibitemOpen
  \bibfield  {author} {\bibinfo {author} {\bibfnamefont {B.~V.}\ \bibnamefont
  {Chirikov}},\ }\href {\doibase DOI: 10.1016/0370-1573(79)90023-1} {\bibfield
  {journal} {\bibinfo  {journal} {Physics Reports}\ }\textbf {\bibinfo {volume}
  {52}},\ \bibinfo {pages} {263 } (\bibinfo {year} {1979})}\BibitemShut
  {NoStop}%
\bibitem [{\citenamefont {Shepelyansky}(1987)}]{shepelyanskyqcorr}%
  \BibitemOpen
  \bibfield  {author} {\bibinfo {author} {\bibfnamefont {D.}~\bibnamefont
  {Shepelyansky}},\ }\href@noop {} {\bibfield  {journal} {\bibinfo  {journal}
  {Physica D}\ }\textbf {\bibinfo {volume} {28}},\ \bibinfo {pages} {103}
  (\bibinfo {year} {1987})}\BibitemShut {NoStop}%
\bibitem [{\citenamefont {Ramareddy}\ \emph {et~al.}(2010)\citenamefont
  {Ramareddy}, \citenamefont {Behinaein}, \citenamefont {Talukdar},
  \citenamefont {Ahmadi},\ and\ \citenamefont {Summy}}]{gil2010b}%
  \BibitemOpen
  \bibfield  {author} {\bibinfo {author} {\bibfnamefont {V.}~\bibnamefont
  {Ramareddy}}, \bibinfo {author} {\bibfnamefont {G.}~\bibnamefont
  {Behinaein}}, \bibinfo {author} {\bibfnamefont {I.}~\bibnamefont {Talukdar}},
  \bibinfo {author} {\bibfnamefont {P.}~\bibnamefont {Ahmadi}}, \ and\ \bibinfo
  {author} {\bibfnamefont {G.~S.}\ \bibnamefont {Summy}},\ }\href
  {http://stacks.iop.org/0295-5075/89/i=3/a=33001} {\bibfield  {journal}
  {\bibinfo  {journal} {EPL (Europhysics Letters)}\ }\textbf {\bibinfo {volume}
  {89}},\ \bibinfo {pages} {33001} (\bibinfo {year} {2010})}\BibitemShut
  {NoStop}%
\bibitem [{\citenamefont {Shrestha}\ \emph {et~al.}(2012)\citenamefont
  {Shrestha}, \citenamefont {Ni}, \citenamefont {Lam}, \citenamefont
  {Wimberger},\ and\ \citenamefont {Summy}}]{gil2012}%
  \BibitemOpen
  \bibfield  {author} {\bibinfo {author} {\bibfnamefont {R.~K.}\ \bibnamefont
  {Shrestha}}, \bibinfo {author} {\bibfnamefont {J.}~\bibnamefont {Ni}},
  \bibinfo {author} {\bibfnamefont {W.~K.}\ \bibnamefont {Lam}}, \bibinfo
  {author} {\bibfnamefont {S.}~\bibnamefont {Wimberger}}, \ and\ \bibinfo
  {author} {\bibfnamefont {G.~S.}\ \bibnamefont {Summy}},\ }\href {\doibase
  10.1103/PhysRevA.86.043617} {\bibfield  {journal} {\bibinfo  {journal} {Phys.
  Rev. A}\ }\textbf {\bibinfo {volume} {86}},\ \bibinfo {pages} {043617}
  (\bibinfo {year} {2012})}\BibitemShut {NoStop}%
\bibitem [{\citenamefont {Shrestha}\ \emph
  {et~al.}(2013{\natexlab{b}})\citenamefont {Shrestha}, \citenamefont {Ni},
  \citenamefont {Lam}, \citenamefont {Summy},\ and\ \citenamefont
  {Wimberger}}]{gil2013b}%
  \BibitemOpen
  \bibfield  {author} {\bibinfo {author} {\bibfnamefont {R.~K.}\ \bibnamefont
  {Shrestha}}, \bibinfo {author} {\bibfnamefont {J.}~\bibnamefont {Ni}},
  \bibinfo {author} {\bibfnamefont {W.~K.}\ \bibnamefont {Lam}}, \bibinfo
  {author} {\bibfnamefont {G.~S.}\ \bibnamefont {Summy}}, \ and\ \bibinfo
  {author} {\bibfnamefont {S.}~\bibnamefont {Wimberger}},\ }\href {\doibase
  10.1103/PhysRevE.88.034901} {\bibfield  {journal} {\bibinfo  {journal} {Phys.
  Rev. E}\ }\textbf {\bibinfo {volume} {88}},\ \bibinfo {pages} {034901}
  (\bibinfo {year} {2013}{\natexlab{b}})}\BibitemShut {NoStop}%
\bibitem [{\citenamefont {d'Arcy}\ \emph {et~al.}(2004)\citenamefont {d'Arcy},
  \citenamefont {Godun}, \citenamefont {Summy}, \citenamefont {Guarneri},
  \citenamefont {Wimberger}, \citenamefont {Fishman},\ and\ \citenamefont
  {Buchleitner}}]{DarcyPRE}%
  \BibitemOpen
  \bibfield  {author} {\bibinfo {author} {\bibfnamefont {M.~B.}\ \bibnamefont
  {d'Arcy}}, \bibinfo {author} {\bibfnamefont {R.~M.}\ \bibnamefont {Godun}},
  \bibinfo {author} {\bibfnamefont {G.~S.}\ \bibnamefont {Summy}}, \bibinfo
  {author} {\bibfnamefont {I.}~\bibnamefont {Guarneri}}, \bibinfo {author}
  {\bibfnamefont {S.}~\bibnamefont {Wimberger}}, \bibinfo {author}
  {\bibfnamefont {S.}~\bibnamefont {Fishman}}, \ and\ \bibinfo {author}
  {\bibfnamefont {A.}~\bibnamefont {Buchleitner}},\ }\href {\doibase
  10.1103/PhysRevE.69.027201} {\bibfield  {journal} {\bibinfo  {journal} {Phys.
  Rev. E}\ }\textbf {\bibinfo {volume} {69}},\ \bibinfo {pages} {027201}
  (\bibinfo {year} {2004})}\BibitemShut {NoStop}%
\bibitem [{\citenamefont {Wimberger}\ \emph {et~al.}(2005)\citenamefont
  {Wimberger}, \citenamefont {Sadgrove}, \citenamefont {Parkins},\ and\
  \citenamefont {Leonhardt}}]{Wimberger2005PRA}%
  \BibitemOpen
  \bibfield  {author} {\bibinfo {author} {\bibfnamefont {S.}~\bibnamefont
  {Wimberger}}, \bibinfo {author} {\bibfnamefont {M.}~\bibnamefont {Sadgrove}},
  \bibinfo {author} {\bibfnamefont {S.}~\bibnamefont {Parkins}}, \ and\
  \bibinfo {author} {\bibfnamefont {R.}~\bibnamefont {Leonhardt}},\ }\href
  {\doibase 10.1103/PhysRevA.71.053404} {\bibfield  {journal} {\bibinfo
  {journal} {Phys. Rev. A}\ }\textbf {\bibinfo {volume} {71}},\ \bibinfo
  {pages} {053404} (\bibinfo {year} {2005})}\BibitemShut {NoStop}%
\bibitem [{\citenamefont {Karski}\ \emph {et~al.}(2009)\citenamefont {Karski},
  \citenamefont {Förster}, \citenamefont {Choi}, \citenamefont {Steffen},
  \citenamefont {Alt}, \citenamefont {Meschede},\ and\ \citenamefont
  {Widera}}]{meschede2009}%
  \BibitemOpen
  \bibfield  {author} {\bibinfo {author} {\bibfnamefont {M.}~\bibnamefont
  {Karski}}, \bibinfo {author} {\bibfnamefont {L.}~\bibnamefont {Förster}},
  \bibinfo {author} {\bibfnamefont {J.-M.}\ \bibnamefont {Choi}}, \bibinfo
  {author} {\bibfnamefont {A.}~\bibnamefont {Steffen}}, \bibinfo {author}
  {\bibfnamefont {W.}~\bibnamefont {Alt}}, \bibinfo {author} {\bibfnamefont
  {D.}~\bibnamefont {Meschede}}, \ and\ \bibinfo {author} {\bibfnamefont
  {A.}~\bibnamefont {Widera}},\ }\href@noop {} {\bibfield  {journal} {\bibinfo
  {journal} {Science}\ }\textbf {\bibinfo {volume} {325}},\ \bibinfo {pages}
  {174} (\bibinfo {year} {2009})}\BibitemShut {NoStop}%
\bibitem [{\citenamefont {Z\"ahringer}\ \emph {et~al.}(2010)\citenamefont
  {Z\"ahringer}, \citenamefont {Kirchmair}, \citenamefont {Gerritsma},
  \citenamefont {Solano}, \citenamefont {Blatt},\ and\ \citenamefont
  {Roos}}]{PhysRevLett.104.100503}%
  \BibitemOpen
  \bibfield  {author} {\bibinfo {author} {\bibfnamefont {F.}~\bibnamefont
  {Z\"ahringer}}, \bibinfo {author} {\bibfnamefont {G.}~\bibnamefont
  {Kirchmair}}, \bibinfo {author} {\bibfnamefont {R.}~\bibnamefont
  {Gerritsma}}, \bibinfo {author} {\bibfnamefont {E.}~\bibnamefont {Solano}},
  \bibinfo {author} {\bibfnamefont {R.}~\bibnamefont {Blatt}}, \ and\ \bibinfo
  {author} {\bibfnamefont {C.~F.}\ \bibnamefont {Roos}},\ }\href {\doibase
  10.1103/PhysRevLett.104.100503} {\bibfield  {journal} {\bibinfo  {journal}
  {Phys. Rev. Lett.}\ }\textbf {\bibinfo {volume} {104}},\ \bibinfo {pages}
  {100503} (\bibinfo {year} {2010})}\BibitemShut {NoStop}%
\bibitem [{\citenamefont {Broome}\ \emph {et~al.}(2010)\citenamefont {Broome},
  \citenamefont {Fedrizzi}, \citenamefont {Lanyon}, \citenamefont {Kassal},
  \citenamefont {Aspuru-Guzik},\ and\ \citenamefont
  {White}}]{PhysRevLett.104.153602}%
  \BibitemOpen
  \bibfield  {author} {\bibinfo {author} {\bibfnamefont {M.~A.}\ \bibnamefont
  {Broome}}, \bibinfo {author} {\bibfnamefont {A.}~\bibnamefont {Fedrizzi}},
  \bibinfo {author} {\bibfnamefont {B.~P.}\ \bibnamefont {Lanyon}}, \bibinfo
  {author} {\bibfnamefont {I.}~\bibnamefont {Kassal}}, \bibinfo {author}
  {\bibfnamefont {A.}~\bibnamefont {Aspuru-Guzik}}, \ and\ \bibinfo {author}
  {\bibfnamefont {A.~G.}\ \bibnamefont {White}},\ }\href {\doibase
  10.1103/PhysRevLett.104.153602} {\bibfield  {journal} {\bibinfo  {journal}
  {Phys. Rev. Lett.}\ }\textbf {\bibinfo {volume} {104}},\ \bibinfo {pages}
  {153602} (\bibinfo {year} {2010})}\BibitemShut {NoStop}%
\bibitem [{\citenamefont {Preiss}\ \emph {et~al.}(2015)\citenamefont {Preiss},
  \citenamefont {Ma}, \citenamefont {Tai}, \citenamefont {Lukin}, \citenamefont
  {Rispoli}, \citenamefont {Zupancic}, \citenamefont {Lahini}, \citenamefont
  {Islam},\ and\ \citenamefont {Greiner}}]{Preiss2015}%
  \BibitemOpen
  \bibfield  {author} {\bibinfo {author} {\bibfnamefont {P.~M.}\ \bibnamefont
  {Preiss}}, \bibinfo {author} {\bibfnamefont {R.}~\bibnamefont {Ma}}, \bibinfo
  {author} {\bibfnamefont {M.~E.}\ \bibnamefont {Tai}}, \bibinfo {author}
  {\bibfnamefont {A.}~\bibnamefont {Lukin}}, \bibinfo {author} {\bibfnamefont
  {M.}~\bibnamefont {Rispoli}}, \bibinfo {author} {\bibfnamefont
  {P.}~\bibnamefont {Zupancic}}, \bibinfo {author} {\bibfnamefont
  {Y.}~\bibnamefont {Lahini}}, \bibinfo {author} {\bibfnamefont
  {R.}~\bibnamefont {Islam}}, \ and\ \bibinfo {author} {\bibfnamefont
  {M.}~\bibnamefont {Greiner}},\ }\href {\doibase 10.1126/science.1260364}
  {\bibfield  {journal} {\bibinfo  {journal} {Science}\ }\textbf {\bibinfo
  {volume} {347}},\ \bibinfo {pages} {1229} (\bibinfo {year}
  {2015})}\BibitemShut {NoStop}%
\bibitem [{\citenamefont {M\"ulken}\ \emph {et~al.}(2007)\citenamefont
  {M\"ulken}, \citenamefont {Blumen}, \citenamefont {Amthor}, \citenamefont
  {Giese}, \citenamefont {Reetz-Lamour},\ and\ \citenamefont
  {Weidem\"uller}}]{PhysRevLett.99.090601}%
  \BibitemOpen
  \bibfield  {author} {\bibinfo {author} {\bibfnamefont {O.}~\bibnamefont
  {M\"ulken}}, \bibinfo {author} {\bibfnamefont {A.}~\bibnamefont {Blumen}},
  \bibinfo {author} {\bibfnamefont {T.}~\bibnamefont {Amthor}}, \bibinfo
  {author} {\bibfnamefont {C.}~\bibnamefont {Giese}}, \bibinfo {author}
  {\bibfnamefont {M.}~\bibnamefont {Reetz-Lamour}}, \ and\ \bibinfo {author}
  {\bibfnamefont {M.}~\bibnamefont {Weidem\"uller}},\ }\href {\doibase
  10.1103/PhysRevLett.99.090601} {\bibfield  {journal} {\bibinfo  {journal}
  {Phys. Rev. Lett.}\ }\textbf {\bibinfo {volume} {99}},\ \bibinfo {pages}
  {090601} (\bibinfo {year} {2007})}\BibitemShut {NoStop}%
\bibitem [{\citenamefont {Xue}\ and\ \citenamefont
  {Sanders}(2013)}]{PhysRevA.87.022334}%
  \BibitemOpen
  \bibfield  {author} {\bibinfo {author} {\bibfnamefont {P.}~\bibnamefont
  {Xue}}\ and\ \bibinfo {author} {\bibfnamefont {B.~C.}\ \bibnamefont
  {Sanders}},\ }\href {\doibase 10.1103/PhysRevA.87.022334} {\bibfield
  {journal} {\bibinfo  {journal} {Phys. Rev. A}\ }\textbf {\bibinfo {volume}
  {87}},\ \bibinfo {pages} {022334} (\bibinfo {year} {2013})}\BibitemShut
  {NoStop}%
\end{thebibliography}

\end{document}